\documentclass[sigconf]{acmart}
\AtBeginDocument{%
  }

\usepackage{multirow}
\usepackage{soul}   
\usepackage{CJKutf8}
\soulregister\cite7

\copyrightyear{2026}
\acmYear{2026}
\setcopyright{cc}
\setcctype{by-nc-nd}
\acmConference[CHI '26]{Proceedings of the 2026 CHI Conference on Human Factors in Computing Systems}{April 13--17, 2026}{Barcelona, Spain}
\acmBooktitle{Proceedings of the 2026 CHI Conference on Human Factors in Computing Systems (CHI '26), April 13--17, 2026, Barcelona, Spain}
\acmPrice{}
\acmDOI{10.1145/3772318.3791771}
\acmISBN{979-8-4007-2278-3/2026/04}

\begin{document}

\title{SparkTales: Facilitating Cross-Language Collaborative Storytelling through Coordinator-AI Collaboration}

\author{Wenxin Zhao}
\email{zhaowx21@m.fudan.edu.cn}
\affiliation{%
  \institution{Fudan University}
  \city{Shanghai}
  \country{China}
}

\author{Peng Zhang}
\authornote{Corresponding authors.}
\email{zhangpeng_@fudan.edu.cn}
\affiliation{%
  \institution{Fudan University}
  \city{Shanghai}
  \country{China}
}

\author{Hansu Gu}
\email{hansug@acm.org}
\affiliation{%
  \institution{Independent}
  \city{Seattle, Washington}
  \country{USA}
}

\author{Haoxuan Zhou}
\email{24210240428@m.fudan.edu.cn}
\affiliation{%
  \institution{Fudan University}
  \city{Shanghai}
  \country{China}
}

\author{Xiaojie Huo}
\email{xh638@stern.nyu.edu}
\affiliation{%
  \institution{Jiedou Edtech, Inc}
  \city{Newport Beach, California}
  \country{USA}
}

\author{Lin Wang}
\email{linw@fudan.edu.cn}
\affiliation{%
  \institution{Fudan University}
  \city{Shanghai}
  \country{China}
}

\author{Wen Zheng}
\email{zhengwen@fudan.edu.cn}
\affiliation{%
  \institution{Fudan University}
  \city{Shanghai}
  \country{China}
}

\author{Tun Lu}
\authornotemark[1]
\email{lutun@fudan.edu.cn}
\affiliation{%
  \institution{Fudan University}
  \city{Shanghai}
  \country{China}
}

\author{Ning Gu}
\email{ninggu@fudan.edu.cn}
\affiliation{%
  \institution{Fudan University}
  \city{Shanghai}
  \country{China}
}

\renewcommand{\shortauthors}{Trovato et al.}

\begin{abstract}

Cross-language collaborative storytelling plays a vital role in children's language learning and cultural development, fostering both expressive ability and intercultural awareness. Yet, in practice, children's participation is often shallow, and facilitating such sessions places heavy cognitive and organizational burdens on coordinators, who must coordinate language support, maintain children's engagement, and navigate cultural differences. To address these challenges, we conducted a formative study with coordinators to identify their needs and pain points, which guided the design of SparkTales, an intelligent support system for cross-language collaborative storytelling. SparkTales leverages both individual and common characteristics of participating children to provide coordinators with story frameworks, diverse questions, and comprehension-oriented materials, aiming to reduce coordinators' workload while enhancing children's interactive engagement. Evaluation results show that SparkTales not only significantly increases coordinators' efficiency and quality of guidance but also improves children's participation, providing valuable insights for the design of future intelligent systems supporting cross-language collaboration.

\end{abstract}




\maketitle

\section{Introduction}
\label{Introduction}

\begin{figure*}[t]
    \begin{minipage}[t]{0.47\linewidth}
        \centering
        \includegraphics[width=\textwidth]{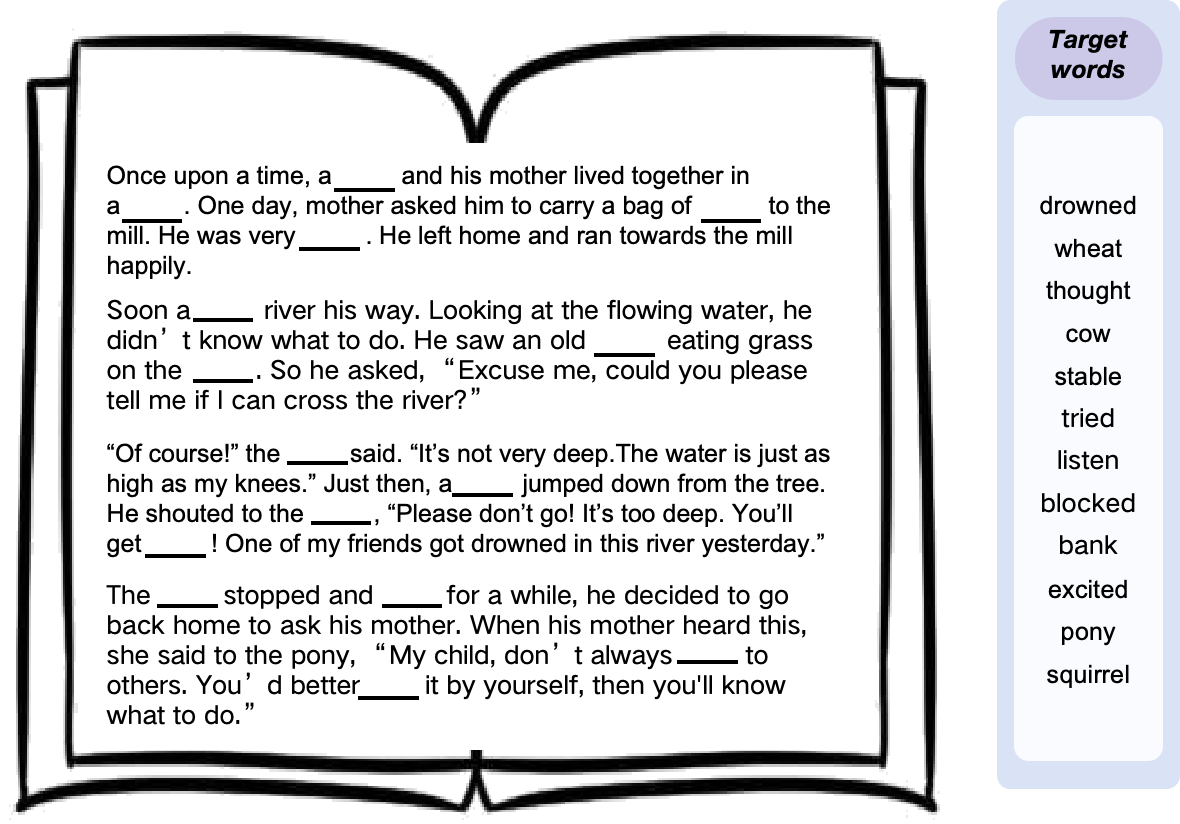}
        \centerline{\small{(a) Monolingual collaborative storybook (English).}}
        \label{fig: Monolingual Collaborative Storybook}
    \end{minipage}
    \hspace{0.5cm}
    \begin{minipage}[t]{0.47\linewidth}
        \centering
        \includegraphics[width=\textwidth]{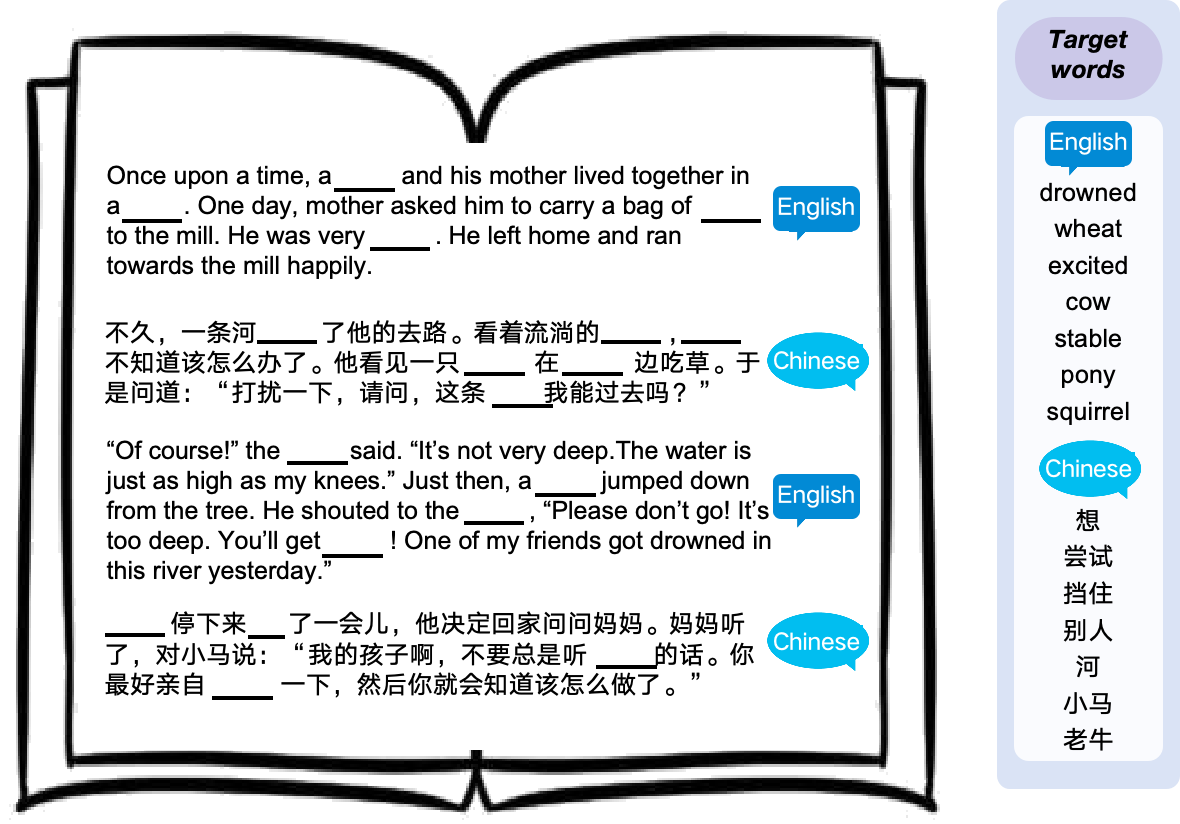}
        \centerline{\small{(b) Bilingual collaborative storybook (English/Chinese).}}
        \label{fig: Bilingual Collaborative Storybook}
    \end{minipage}
    \caption{Examples of monolingual and bilingual collaborative storybooks.}
    \label{fig:monolingual and bilingual collaborative storybooks}
\end{figure*}

Collaborative storytelling has been widely adopted in children's language and cultural learning, serving as a method to enhance children's language expression and comprehension \cite{ryokai1999storymat, shakeri2021saga}.
\textcolor{black}{In this practice, a teacher or parent typically acts as a coordinator, providing target vocabulary (monolingual or bilingual) and a basic story framework. They guide children to use the target language to elaborate on details, adapt the narrative, and extend the story through questioning and feedback \cite{wang2003collaborative, Dressel_2021}.}

To maintain a balance between storytelling efficiency and quality, collaborative storytelling is often conducted in pairs, with one coordinator working with two children 
\textcolor{black}{who take turns contributing to the storyline \cite{swain1998interaction, storch2002patterns}, resulting in the generation of a monolingual or bilingual collaborative storybook, as illustrated in Figure~\ref{fig:monolingual and bilingual collaborative storybooks}.}
\textcolor{black}{In these scenarios, collaborative storytelling serves as a key approach to linguistic and intercultural development. It engages children in simulated situations, role-play, and interactive language use, which promotes cognitive reflection and the flexible application of linguistic and cultural knowledge \cite{kirsch2016developing, nishioka2016analysing, zalka2017collaborative}.}
\textcolor{black}{Moreover, alternating between Storyteller and Storylistener roles, where one extends the narrative while the other interprets it, supports collaborative development through interaction and negotiation. This dynamic fosters reciprocal peer teaching, authentic language use, and proactive participation in paired learning \cite{alfatihah2022teaching}.}

Despite these advantages, cross-language collaborative storytelling faces new challenges, among which insufficient engagement is particularly notable \cite{huang2017student}. 
\textcolor{black}{Cross-language collaborative storytelling primarily targets children aged 7–11, who fall into Piaget's concrete operational stage of cognitive development \cite{Ummah2018, ramadhania2020paired}. At this stage, their developing abstract reasoning capabilities can lead to pragmatic misunderstandings, topic breakdowns, and limited interaction strategies during language exchanges \cite{piaget2013construction, huitt2003piaget}.}
Cultural differences add further complexity, creating obstacles in interpreting peers' intentions and reducing willingness to communicate \cite{gallagher2013willingness}. 
These factors mean that successful cross-language collaborative storytelling depends heavily on the guidance of a coordinator \cite{toth2008teacher}. 
However, facilitation is demanding: coordinators must juggle real-time tasks such as topic steering, vocabulary integration, and question design, all of which require considerable knowledge and skill \cite{strati2017perceived}. 
At the same time, cultural gaps and generational differences in both coordinator–child and child–child interactions can cause misinterpretations and communication barriers \cite{maunsell2020dyslexia}. 
Given these constraints, teachers and parents often struggle to maintain effective facilitation, leading to reduced children's engagement in collaborative storytelling \cite{strati2017perceived, wong2017too}. 
These challenges motivate us to explore how intelligent systems might support coordinators in reducing workload while sustaining children's engagement. 
Importantly, due to the risks inherent in children's educational contexts, such as concerns around appropriateness, cultural sensitivity, and online safety, the role of adult coordinators remains essential. 
Therefore, the tool we design is not intended to replace coordinators but to serve as an assistant supporting their role.

\textcolor{black}{Large Language Models (LLMs) offer a promising technical foundation for developing intelligent assistance in collaborative storytelling. Leveraging their extensive knowledge and generative capabilities, LLMs can support coordinators by suggesting questions, offering real-time feedback, and assisting with narrative adaptation when specific themes or vocabulary prove challenging \cite{chen-etal-2017-reading, karpukhin2020dense}.
Furthermore, LLMs demonstrate significant cross-cultural capabilities: in multilingual contexts, they can identify cultural nuances and generate responses that are semantically and emotionally aligned with the intended cultural background \cite{conneau2019unsupervised, li2024culture}.
Finally, the flexible conversational nature of LLMs facilitates highly interactive and engaging experiences \cite{billings2024using, khosravi2022explainable}.
}

\textcolor{black}{Nevertheless, designing an LLM-based assistant for collaborative storytelling presents several unresolved challenges.
First, the specific roles and functions of such a system remain ill-defined. There is often uncertainty regarding the timing, modality, and form of support required, complicating the design of appropriate functional modules.
Second, the system must balance competing objectives: it needs to reduce the coordinators' cognitive and operational workload while simultaneously sustaining children's engagement and willingness to express themselves across diverse cultural contexts—a balance that current LLMs often struggle to achieve \cite{li2025demod, liu2025large}.
Third, the complex tasks involved in collaborative storytelling, such as vocabulary integration, question design, and resource preparation, impose high demands on the cross-cultural generation capabilities of LLMs.
Finally, evaluating these systems in real-world scenarios is difficult due to dynamic contexts and multiple stakeholders, necessitating multidimensional metrics such as coordination efficiency and children engagement levels.}

\textcolor{black}{To address these challenges, this paper focuses on a teacher-coordinated cross-language collaborative storytelling scenario where a teacher facilitates interaction between two children—one learning Chinese and the other English.
We initially conducted a formative study with 10 experienced teachers to identify their needs and expectations for an intelligent assistant. From these insights, we derived four key design goals to support coordinators in managing collaborative activities.
Guided by these goals, we developed SparkTales, an intelligent assistant designed to support coordinators in online cross-language collaborative storytelling. The system helps navigate multifaceted tasks and cultural challenges while fostering deeper interaction and active participation among children.
SparkTales integrates five core modules: a Configuration Module for specifying target vocabulary and children profiles; an Individual Characteristic Summarization Module; a Common Characteristic Summarization Module; a Collaborative Storytelling Support Module, which generates story frameworks, diverse questions, and comprehension materials; and a Review and Feedback Module that analyzes engagement patterns and suggests configuration updates.
We implemented SparkTales and conducted an evaluation with 8 experienced teachers and 8 pairs of children. The results demonstrate that SparkTales enhanced both coordination efficiency and facilitation quality, as well as children's interactive engagement, eliciting strong interest in continued use from participants.}

The main contributions of this study can be summarized as follows:
\begin{itemize}
    \item We present the first study of an assistant system designed specifically for coordinators in collaborative storytelling, aiming to reduce workload while enhancing children's engagement.
    
    \item Through a formative study, we systematically identify the core needs and expectations of teachers in coordinating cross-language collaborative storytelling.
    
    \item We design and implement SparkTales, an intelligent assistant for coordinator-supported storytelling, and evaluate its effectiveness in improving children's engagement and facilitator support.
    
    \item We propose new insights to guide the design of future intelligent systems that assist coordinators in educational contexts.
    
\end{itemize}


The remainder of this paper is structured as follows. Section~\ref{Related Works} reviews related work. Section~\ref{Formative Study} describes our formative study. Section~\ref{SparkTales} introduces the SparkTales framework and implementation. Section~\ref{Evaluation} presents our evaluation methods and results. Section~\ref{Discussion} analyzes key findings. Section~\ref{Ethical Consideration} and Section~\ref{Limitations and Future Work} discuss ethics and limitations, and Section~\ref{Conclusion} concludes. Finally, Section~\ref{Acknowledgments of the Use of AI} explains the use of AI in this paper.


\section{Related Work}
\label{Related Works}

\subsection{(Collaborative) Storytelling}

Storytelling is widely used to promote children's language development, cognitive skills, and emotional expression \cite{mercer2007dialogue}. Bruner noted that storytelling can not only help children construct meaning and comprehend complex situations but also develop their thinking skills \cite{bruner2009actual}. Advancements in digital technology have transformed storytelling into a multimodal and integrative activity \cite{robin2012educators}. With multimodal presentation and cultural integration, digital storytelling enhances children's language development, literacy, and cultural identity \cite{robin2016power}. Empirical evidence further demonstrates that it can improve children's oral complexity, narrative coherence, and comprehension, highlighting its effectiveness for language learning \cite{isbell2004effects}. 

Currently, AI-powered story generation tools and conversational agents are broadening research perspectives on storytelling \cite{li2024we}. For example, Wang and Kreminski demonstrated that incorporating Answer Set Programming (ASP) techniques can effectively guide and enrich LLM–based story generation, thereby enhancing both content diversity and coherence \cite{wang2024guiding}. The Storybuddy enables parent–child to co-reading stories through a chatbot, offering flexible parental guidance to promote children's language development \cite{zhang2022storybuddy}. Similarly, StoryDrawer integrates AI with interactive drawing to enhance children's visual narrative creativity \cite{zhang2022storydrawer}. Furthermore, \cite{li2024we} examined storytelling from a human–AI collaboration perspective, highlighting AI's role in enriching interactions and informing design for children's storytelling. In addition, open datasets related to storytelling, such as FairytaleQA \cite{fantastic} and StorySparkQA \cite{chen2023storysparkqa}, provide rich resources and evaluation foundations for research on story understanding and Question Answering (QA) generation.

\textcolor{black}{Collaborative storytelling, wherein children co-create stories with peers through activities such as simulated scenarios, role-playing, and reciprocal questioning, emphasizes peer interaction and cooperation \cite{zalka2017collaborative}.
As illustrated in Figure~\ref{fig:monolingual and bilingual collaborative storybooks}, in a monolingual context, both children learn the same language and take turns contributing, ultimately producing a monolingual collaborative storybook (Figure~\ref{fig:monolingual and bilingual collaborative storybooks} (a)).
In contrast, in a bilingual context involving one child learning Chinese and another learning English, participants alternate between the two languages to adapt or expand the story. This process generates a bilingual collaborative storybook featuring alternating Chinese and English paragraphs (Figure~\ref{fig:monolingual and bilingual collaborative storybooks} (b)).
This structured practice stimulates verbal interaction and narrative skills, enhances social and problem-solving abilities, and fosters intercultural awareness, thereby facilitating effective language acquisition and cross-cultural communication competence \cite{colas2017interaction, piipponen2019children}.}

Despite the educational value, research on collaborative storytelling - particularly in online cross-language environments - remains limited within the Human-Computer Interaction (HCI) field, with most existing studies focusing on physical, face-to-face collaboration or digital co-creative platforms. For example, KidPad promotes social learning through role allocation and turn-taking, allowing multiple children to draw and link story scenes in a 2-dimensional space \cite{hourcade2002kidpad}. The ShadowStory project incorporates cultural heritage into a multi-participant digital storytelling platform, facilitating cultural transmission and collaborative creation among diverse participants \cite{lu2011shadowstory}. Additionally, some storytelling technologies specifically designed for young children emphasize simple and intuitive interaction to support adult-guided collaborative activities \cite{benford2000designing}. By leveraging LLMs, SAGA enables collaborative storytelling in which an AI agent and a human take turns adding content to a story \cite{nichols2020collaborative}. Although these studies offer diverse support for collaborative storytelling, systematic exploration and in-depth empirical research on children's collaboration in online cross-language contexts remain limited.

\subsection{\textcolor{black}{Coordinator–AI Teaming for Child Interaction and Collaboration}}
\label{Related Works: Coordinator–AI Teaming for Child Interaction}

\textcolor{black}{In recent years, employing AI to support children's learning, creativity, and collaboration has become a significant trajectory in HCI research \cite{tolksdorf2021comparing,smakman2022robotic}.
In this context, an increasing number of studies emphasize that AI is no longer merely a tool, but an active partner in interacting and collaborating with children \cite{sievers2025practical,abbasi2025longitudinal,rudenko2024child}.
Concurrently, research has begun to explore how peer interaction and collaboration can be transformed through AI integration.
In games or creative activities, AI systems are now embedded into children's contexts as mediators, guides, or partners.
These systems not only facilitate knowledge sharing and task division but also enhance children's social skills, expressive initiative, and creativity through dynamic processes such as verbal communication, role-playing, and content generation \cite{luckin2016intelligence}.}

\textcolor{black}{However, despite these advantages, studies highlight the cognitive, ethical, and safety risks children face when interacting directly with intelligent systems \cite{jiao2025llms,kurian2024no}.
Therefore, the involvement of an adult coordinator (e.g., a teacher or parent) is essential, necessitating a Human-AI Teaming model to support children's social activities.
In this model, the coordinator and AI establish clear task divisions, maintain information-sharing mechanisms, and adapt dynamically to contextual changes \cite{wang2025adaptive, hollan2000distributed}.
Specifically, the coordinator generally supervises children's understanding and execution of tasks, while the AI continuously collects and analyzes behavioral data \cite{trajkovski2025ai}.
Through this collaborative approach, the cognitive process is distributed across the coordinator, the AI, and the interaction environment (including task materials and real-time behaviors), forming a Distributed Cognition (DCog) system \cite{hollan2000distributed}.
This distributed structure enables information to be externalized via AI-generated content, shared with children through coordinator-mediated presentation, and continuously refined as both the coordinator and AI adapt to the children's real-time responses \cite{sidji2024adopting}.
By allocating cognitive responsibilities in this manner, the DCog system enhances the overall collaboration efficiency and decision-making quality of the Coordinator–AI team, while simultaneously maintaining safety and pedagogical professionalism \cite{belpaeme2018social}.}

\textcolor{black}{Building on this allocation, coordinators and AI can adopt different collaboration strategies depending on the context, resulting in varying levels of AI visibility to children.
In the first mode, the coordinator actively leads children in interacting with the AI to accomplish tasks. Here, the AI is \textit{visible}: children perceive and respond to AI-generated content directly, but their interaction remains scaffolded by the coordinator to ensure task goals and safety are met \cite{ekstrom2025teaching, ekstrom2022dual}.
Conversely, in the second mode, the coordinator acts as the sole guide, while the AI functions as a backend assistant providing strategic suggestions and feedback. In this scenario, the AI is \textit{invisible} to the children; they interact exclusively with the coordinator, benefiting from AI support while remaining shielded from direct system interaction \cite{ceha2022identifying,ekstrom2025teaching}.
These collaboration modes reflect the complementary roles of the coordinator and AI in guiding children's activities, providing theoretical support for conducting our research.}

\subsection{Technology-Supported Cross-Language Learning}

With increasing cultural exchanges under globalization, cross-language learning has become a focus in educational technology design, where both educators and learners face multifaceted challenges, including linguistic differences and cultural misunderstandings \cite{gunawardena2018culturally}. 
To achieve equity and inclusion in cross-language environments, educators must flexibly adapt teaching strategies to learners' language and cultural backgrounds, balancing interaction efficiency and enhancing cultural sensitivity \cite{gay2018culturally}. 
Recent advances in HCI have introduced innovative technologies such as visual interfaces, multimodal context-aware systems, and dynamic feedback mechanisms into cross-language learning, facilitating teacher-student understanding and communication \cite{morales2020nationality, li2024addressing}. For instance, Edge et al. proposed the MicroMandarin mobile learning system, which employs environmental sensors to dynamically adjust instruction, enhancing situational authenticity and learner engagement \cite{edge2011micromandarin}. 
The emergence of AI-Generated Content (AIGC) has further advanced personalized and multimodal approaches to cross-language learning. By generating multimodal resources tailored to learners' profiles, AIGC significantly enhances cross-language learning efficiency while improving the cultural relevance and personalization of learning materials \cite{AIGC_Technology}. For instance, Yan et al. demonstrated ChatGPT's effectiveness in L2 writing productivity and provided practical implications for AI-assisted language pedagogy \cite{yan2023impact}, while Yang et al. validated that culturally adapted AIGC-generated visuals can improve instructional efficiency \cite{yang2023depth}.

Paired learning in the cross-language background is an instructional strategy that emphasizes equal dialogue and knowledge co-construction, demonstrating significant advantages in enhancing targeted mutual support \cite{o2020intercultural}. In this setting, two students from different linguistic and cultural backgrounds engage in peer-to-peer learning, which can be either coordinated by a coordinator (e.g., a teacher) or entirely self-organized \cite{kohn2011web}, such as an American learning Chinese collaborating with a Chinese learning English through interactive activities like collaborative storytelling. In this process, learners adapt to different linguistic structures and cultural contexts, fostering effective communication and deeper intercultural understanding \cite{kohn2011web}. For instance, Alanís and Arreguín-Anderson emphasized that paired learning can stimulate positive interaction and responsibility among learners, improving the depth and quality of collaborative learning \cite{alanis2019paired}; Huseynli noted that structured pairing strategies boost learners' language output and develop critical thinking skills \cite{huseynli2024power}.
With advances in digital media and AI, paired learning models increasingly leverage technology to enhance cross-linguistic learning, providing more personalized, immersive participation and collaboration \cite{Ummah2018}. For example, Ummah et al. found that incorporating digital storytelling media into paired learning enhanced students' participation while fostering comprehension of story content and intercultural awareness \cite{Ummah2018}.

In summary, designing technologies and systems to support collaborative storytelling has remained an important research area in HCI. Collaborative storytelling, as an essential activity through peer interaction, can promote both children's language development and cross-cultural understanding, while existing HCI research in this area remains limited. With the rapid advancement of AI, particularly AIGC, AI-powered cross-language learning has attracted multidisciplinary attention. However, cross-language collaborative storytelling introduces new challenges due to cultural differences and complex collaboration requirements. Therefore, this paper investigates how AI technologies can effectively support collaborative storytelling in cross-language learning contexts, aiming to optimize coordinator guidance and enhance children's engagement.

\section{Formative Study}
\label{Formative Study}

Cross-language collaborative storytelling among children is a special collaboration practice without sufficient investigation. To inform our design, we first conducted a formative study to explore coordinators' current practices, challenges faced, and the corresponding expectations in this context.


\subsection{Method}
\subsubsection{Participants}
\label{formative_Participants}

This study focused on teacher-coordinated collaborative storytelling, involving one teacher and two children (learning Chinese and English, respectively), to investigate tools that support coordinators in such a context. We recruited teacher participants through online and offline surveys, with the following inclusion criteria: (1) having practical experience in cross-language collaborative storytelling activities for primary school children (approximately aged 7–11), and (2) demonstrating an open attitude toward AI-assisted tools. Through the snowball sampling method \cite{naderifar2017snowball}, we selected 10 teachers for this study. The demographic characteristics of the participants are shown in Table \ref{tab:Summary of participants}, where T represents teachers. Given that language teaching and collaborative storytelling are the fundamental context of our study, we specifically focused on two key characteristics of participants: teachers' years of language teaching and frequency of conducting collaborative storytelling.


\begin{table*}[ht] 
\small
\caption{Demographics of formative study participants.}
\label{tab:Summary of participants}
\centering
\resizebox{\linewidth}{!}{
\begin{tabular}{c|cccccccc}
\hline
\textbf{ID}   & \textbf{Age} & \textbf{Gender} & \textbf{Education} &
\textbf{Major}&
\textbf{Age Group Taught}&
\textbf{Language Taught}&
\textbf{\begin{tabular}[c]{@{}c@{}}Years of \\ Language Teaching\end{tabular}}&
\textbf{\begin{tabular}[c]{@{}c@{}}Frequency of  \\Collaborative Storytelling\end{tabular}} \\ 
\midrule
\textbf{T1}  & 35-44                                         & Male            & Master's Degree                              & Business, Education      &  \begin{tabular}[c]{@{}c@{}}Primary School,\\ Middle School\end{tabular}  &  English, Chinese &   6–10 years                           & 3 or more times per week \\
\midrule
\textbf{T2}  & 35-44                                        & Female            & Master's Degree                              & English Education    & Primary School   &  English, Chinese & 6–10 years & 1–2 times per week\\
\midrule
\textbf{T3}  & 35-44                                        & Male            & Bachelor's Degree                           & Education    & \begin{tabular}[c]{@{}c@{}}Primary School,\\ Middle School\end{tabular}   &  English, Chinese & 6–10 years                           & 3 or more times per week\\
\midrule
\textbf{T4}  & 35-44                                        & Female            & Master's Degree                              & English Education    & \begin{tabular}[c]{@{}c@{}}Preschool, \\Primary School\end{tabular}   &  English, Chinese & 6–10 years & 1–2 times per week\\
\midrule
\textbf{T5}  & 25-34                                        & Female            & Master's Degree                              & TCSOL    & \begin{tabular}[c]{@{}c@{}}Primary School,\\ Middle School\end{tabular}   &  English, Chinese & 3-5 years & 1–2 times per week\\
\midrule
\textbf{T6}  & 25-34                                        & Female            & Master's Degree                              & Computer Science    & All Age Groups   &  English, Chinese, French & 6–10 years & 1–2 times per week\\
\midrule
\textbf{T7}  & 25-34                                        & Female            & Master's Degree                              & TCSOL    & All Age Groups   &  English, Chinese & 3-5 years & 1–2 times per week\\
\midrule
\textbf{T8}  & 35-44                                        & Female            & Master's Degree                              & Social Work    & All Age Groups   &  English, Chinese & 6–10 years & About once a month\\
\midrule
\textbf{T9}  & 35-44                                        & Female            & Master's Degree                              & TCSOL    & \begin{tabular}[c]{@{}c@{}}Primary School,\\ Middle School\end{tabular}   &  English, Chinese & More than 10 years  & 1–2 times per week\\
\midrule
\textbf{T10}  & 25-34                                        & Female            & Master's Degree                              & TCSOL    & All Age Groups   &  English, Chinese & 3-5 years & 3 or more times per week\\ \hline
\end{tabular}
}
\raggedright
\footnotesize
Note: TCSOL = Teaching Chinese to Speakers of Other Languages.
\end{table*}

\subsubsection{Procedure}
\label{formative_procedure}

We conducted semi-structured interviews with these 10 teachers via online meetings, each lasting 40–60 minutes. The interviews focused on three key dimensions. First, we investigated teachers' practical approaches to organizing and guiding children's interactions during online cross-language collaborative storytelling. Second, we explored their challenges and corresponding strategies when conducting collaborative storytelling. Third, we further focused on teachers' expectations for AI-assisted tools. All interviews were recorded and transcribed with participant consent. Participants were assured that all interview-related data and materials would remain confidential. Each participant was compensated with 150 RMB.


For data analysis, we employed the thematic analysis proposed by Braun and Clarke to code the interview transcripts \cite{braun2006using, brule2020thematic}, strictly following the six-phase codebook approach \cite{braun2013successful, braun2014thematic, braun2021one}. The entire process was carried out by three researchers, emphasizing systematization, iteration, and consensus building. All transcripts were first independently reviewed to identify key concepts, themes, and patterns for coding, followed by open coding to extract initial codes on coordination strategies, children's engagement, and technology use. Through iterative discussions, codes were consolidated into higher-order themes, collaboratively refined, defined with clear terms and examples, and finally applied across the data with representative examples selected to support the findings. Throughout the process, regular discussions and iterative refinements ensured transparency, accuracy, and reliability, and coding concluded once all researchers reached consensus on the final themes.

\subsection{Findings}

\subsubsection{Current Practices}

\begin{figure*}
  \centering
  \includegraphics[width=\linewidth]{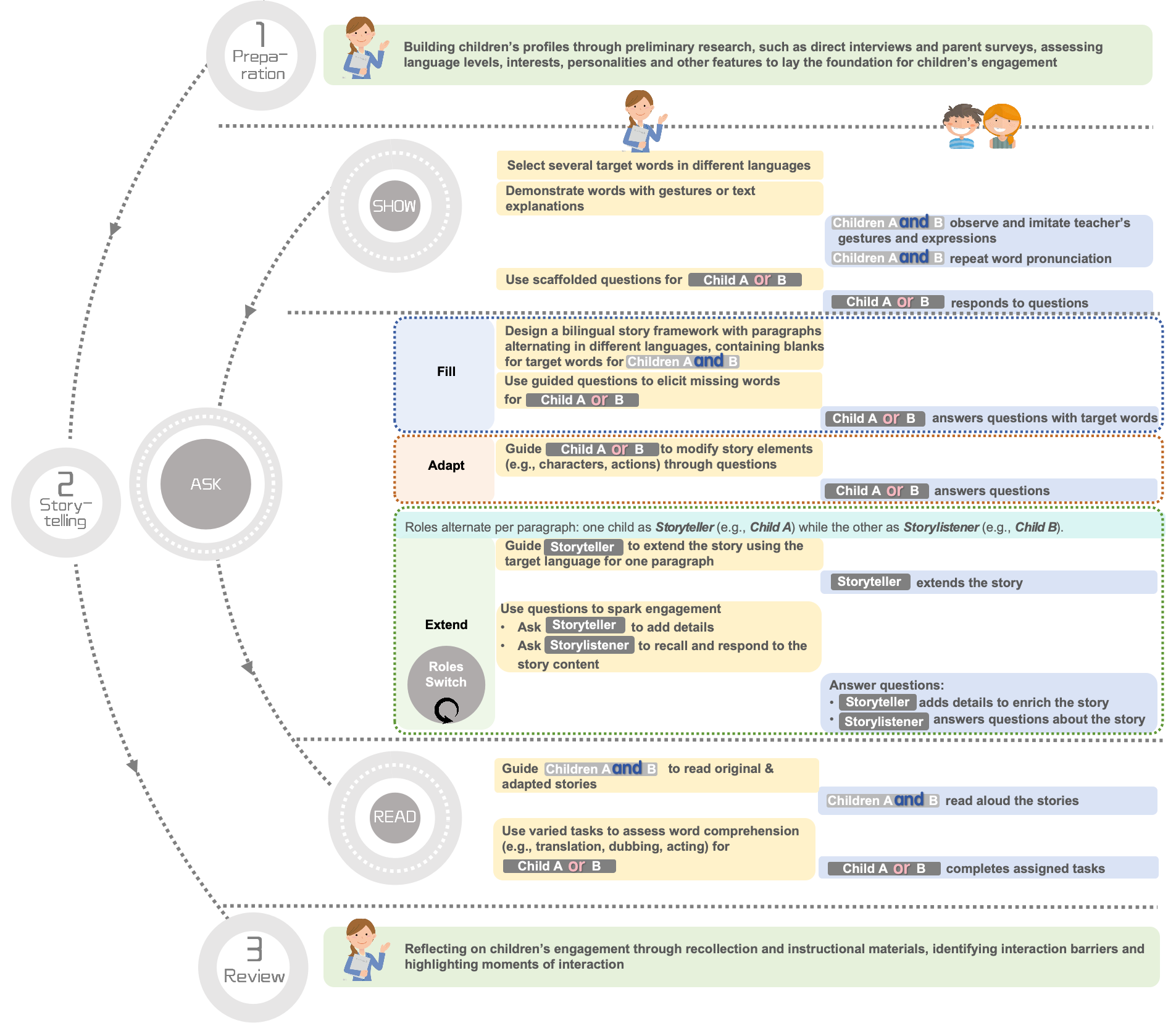}
  \caption{The current process of cross-language collaborative storytelling.}
  \label{fig:storytelling_process}
\end{figure*}

Several advantages of collaborative storytelling, like interactivity, playfulness, and effectiveness, have been highlighted by multiple teachers, for example, T2 described it as "\textit{an effective method to learn language by co-creating stories}". Furthermore, teachers typically structure cross-language collaborative storytelling into three phases -  Preparation, Storytelling, and Review - to enhance children's engagement and interactional continuity, as illustrated in Figure \ref{fig:storytelling_process}. In paired scenarios, teachers often try to match children by similar features to improve peer interaction and collaboration \cite{howe2012children, leaper1991influence, belly2024enhancing}.


In Preparation, teachers generally collect children's characteristics to support peer interaction, as T5 uses "\textit{chats or observations}", while T8 emphasizes understanding "\textit{language levels and personalities}" to design appropriate tasks. During Storytelling, teachers facilitate engagement through three structured phases: (1) Vocabulary Introduction (Show): Teachers select target words, for example, T8 "\textit{explains meanings}", while T7 employs "\textit{body language}" to help comprehension\textcolor{black}{, and use scaffolded questions, a series of step-by-step questions that start simply and gradually become more challenging, to support children’s understanding and use of new words.}
(2) Story Co-creation (Ask): Teachers present a bilingual story framework with "\textit{target words as blanks}"  (T10), \textcolor{black}{and children are guided through questioning to complete a story cloze task, inferring and filling in the blanks based on contextual cues and prior knowledge.}
Open-ended questions then encourage to adapt and extend the story, producing narratives unique to the children, as T9 does by "\textit{replacing characters to enrich scenes}". In paired scenarios, the two children alternate as Storyteller and Storylistener; the Storyteller "\textit{extends and elaborates the story through questioning}" (T7), while the Storylistener "\textit{responds to story content questions}" (T7). 
\textcolor{black}{This process incorporates "\textit{roles switching}" (T8), in which children alternate between the roles of Storyteller and Storylistener. It enables each child to continue the story in their target language, finally producing a bilingual storybook with alternating paragraphs.}
(3) Reading Reinforcement (Read): Teachers consolidate participation using strategies such as "\textit{story dubbing or retelling}" (T9). Finally, in the Review stage, teachers reflect on interactions and summarize key points for future design, as T2 "\textit{focuses on awkwardness moments and analyzes the reasons}".

\subsubsection{Key Findings}
We further analyzed and summarized the prominent challenges these teachers encountered in cross-language storytelling, along with their expectations for intelligent assistance tools during the \textbf{Preparation (F1)}, \textbf{Storytelling (F2, F3, F4)}, and \textbf{Review (F5)} stages.


\textbf{\textcolor{black}{F1: Enable Child Interaction Through Feature Configuration and Matching.}}
Teachers commonly observed that communication barriers often arise from differences in language proficiency and cultural background, while variations in gender, personality, and interests further affect interaction, as T4 noted, "\textit{differences in features largely determine how proactively children engage}". Furthermore, children's limited understanding of peers often reduces responsiveness, as T2 observed, "\textit{they know nothing about each other}". Accordingly, teachers expected the tool to facilitate feature configuration and matching to promote interactions among children, as T5 suggested "\textit{configuring children's interests to supplement storytelling content}". Furthermore, T8 emphasized the importance of "\textit{summarizing their commonalities and generating follow-up materials}".

\textbf{\textcolor{black}{F2: Generate Story Frameworks and Questions to Facilitate Engagement.}} 
During collaborative storytelling, teachers are generally required to conduct several coordination tasks to encourage children's participation, including constructing story frameworks around target words and designing guiding questions. These tasks significantly depend on coordinators' linguistic knowledge, expertise and pedagogical experience, posing coordination challenges. As T6 noted, "\textit{I sometimes can't come up with suitable questions}", affecting storytelling continuity and interaction. Therefore, teachers expected the intelligent tool to generate story frameworks and diverse questions aligned with storylines and children's features to spark their interest in expression. For example, teachers suggested "\textit{generating several story frameworks based on target vocabulary}" (T6) and "\textit{using diverse questions}" (T7) to guide storytelling.

\textbf{\textcolor{black}{F3: Provide Multimodal Support for Language and Cultural Comprehension.}} 
Several teachers reported difficulties interpreting children's expressions across cultures. As T6 noted, she "\textit{sometimes struggles with unfamiliar cultural references}". Children also experienced mutual comprehension difficulties, often "\textit{not understanding what the other is saying}" (T2), arising at two levels: language, such as unfamiliar vocabulary or expressions ("\textit{some words are new to the children}" (T6)), and culture, including differences in traditions or popular content ("\textit{the Chinese child mentioned Chinese square dancing, which confused the American child}" (T2)). Therefore, teachers highlighted the need for the system to supplement real-time contextual and background knowledge via multimodal materials. For instance, T6 suggested "\textit{giving explanations for unfamiliar words}", while T2 recommended "\textit{supplementing images}".

\textbf{\textcolor{black}{F4: Balance Engagement Through Individual and Common Characteristics.}} 
Due to the openness of storytelling, children's unclear roles and interaction rules often cause uneven speaking, disrupting engagement balance and continuity, as T4 observed "\textit{one child speaks actively while the other has no chance}". Thus, teachers suggested the tool should go beyond basic functionalities to balance engagement by considering both individual and common characteristics among children. For example, story frameworks should "\textit{reflect children's common interests}" (T7) while questions and materials should "\textit{align with each child's preferences}" (T1). Turn rotation and speaking-time monitoring were proposed to prevent domination or "free-riding", as T4 proposed "\textit{clarifying Storyteller and Storylistener roles}".

\textbf{\textcolor{black}{F5: Track Feedback Automatically for Reflection.}}
Although reviewing children's participation helps coordinators identify strengths and areas for improvement ("\textit{which child was more active and which topic was more popular}" (T7)), many teachers struggled to simultaneously guide interactions while recording performance, as T1 noted, "\textit{many details are often forgotten}". Accordingly, teachers expected the tool to automatically help capture children's engagement and provide feedback for reflection and future configuration, with T4 noting "\textit{automatic summarization would save time}" and T5 hoping "\textit{results could inform personalized configurations}".


\subsubsection{Design Goals}
\label{Design Goals}
Based on the findings, we propose the following design goals for an intelligent tool to assist coordinators in collaborative storytelling:

\begin{itemize}
    \item \textbf{D1: Support coordinators in configuring and dynamically updating children's individual and common characteristics (F1).} 
    The tool should support coordinators in configuring target words and children's features before collaborative storytelling, which can be dynamically updated to support story co-creation.

    \item \textbf{D2: Support collaborative storytelling based on both individual and common characteristics.} 
        \begin{itemize}
            \item \textbf{D2-1: Generate story frameworks based on common characteristics and target words (F1, F2, F4).} The system should generate fill-in-the-blank story frameworks by combining target vocabulary with children's common characteristics to promote engagement and facilitate subsequent interaction.

            \item \textbf{D2-2: Generate diverse follow-up questions based on individual characteristics (F1, F2, F4).} The system should generate both targeted and heuristic questions according to each child's characteristics, integrating structured questions that scaffold story elements, as well as explicit and implicit questions to spark imagination and reinforce story understanding.
            
            
            \item \textbf{D2-3: Generate comprehension-oriented multimodal material based on individual characteristics and interaction context (F1, F3, F4).} The system should dynamically provide supportive materials, such as images or text, based on each child's individual characteristics and ongoing interactions, to help address cultural differences and facilitate understanding.
        \end{itemize}
        
    \item \textbf{D3: Provide interaction reviews and feedback based on collaborative storytelling process (F1, F5).} The system should continuously track children's audio, text, and role-based interactions to generate visualized reviews and feedback for coordinators to analyze children's engagement and adjust configurations.
    
    \item \textbf{D4: Ensure coordinator control (F1, F2, F3, F4, F5).} The system should be designed to provide coordinators with flexible control over its assistance, enabling easy and timely interventions while preventing over-automation and out-of-control, thereby ensuring effective coordination and avoiding potential risks to children.

\end{itemize}

\section{SparkTales: Facilitating Cross-Language Collaborative Storytelling through Coordinator-AI Collaboration}
\label{SparkTales}
Based on the design goals, we developed SparkTales, an LLM-based intelligent tool for assisting coordinators in cross-language collaborative storytelling. 
We will detail how SparkTales is designed and implemented.


\subsection{SparkTales Design}

\begin{figure*}[htbp]
  \centering
  \includegraphics[width=\linewidth]{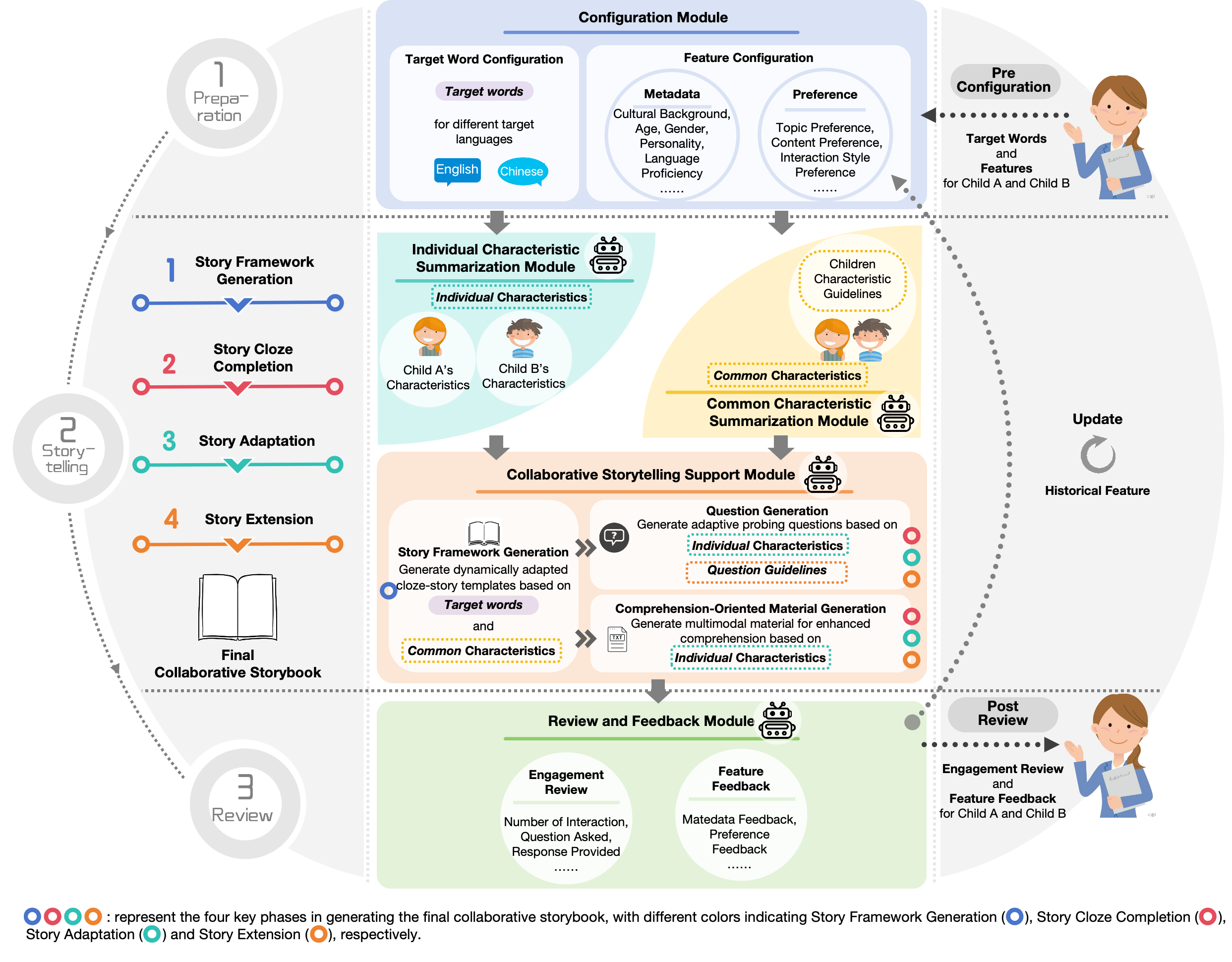}
  \caption{The framework of SparkTales.}
  \label{fig:overview}
\end{figure*}

The overall architecture of SparkTales is illustrated in Figure \ref{fig:overview}, which consists of five core modules:

\begin{itemize}
    \item \textbf{Configuration Module}: This module enables coordinators to select target vocabulary aligned with the goals of the collaborative storytelling and configure each child's features, including Metadata (e.g., cultural background, age, gender, language proficiency) and Preference (e.g., preferred topic, content, interaction style).

    \item \textbf{Individual Characteristic Summarization Module}: Leveraging pre-configured features, this module extracts and summarizes each child's personalized characteristics, resulting in Individual Characteristics for each child that highlight differences in preferences and abilities. These characteristics provide critical support for the subsequent generation of content that emphasizes personal traits.

    \item \textbf{Common Characteristic Summarization Module}: Based on the pre-configured features, this module utilizes the Guideline-driven mechanism to systematically identify shared characteristics among children in terms of interests and other preferences. It generates Common Characteristics that inform the content generation in subsequent sharing tasks.

    \item \textbf{Collaborative Storytelling Support Module}: Using both Individual Characteristics and Common Characteristics, this module generates story frameworks, diverse questions, and comprehension-oriented materials for the collaborative storytelling process. It assists coordinators in guiding children to continuously express themselves and collaborate on tasks - such as filling in gaps, adapting the plot, and extending the story - ultimately producing the complete Final Collaborative Storybook.

    \item \textbf{Review and Feedback Module}: After storytelling, this module automatically generates the engagement review based on children's participation, providing coordinators with a reference to assess the quality of the activity. Additionally, the module also gives suggestions for the existing configured features, facilitating the continuous optimization of collaborative storytelling.

\end{itemize}

The system is mainly powered by LLMs, with all modules except the Configuration Module relying on them, as indicated by the robot icon in Figure \ref{fig:overview}.

\subsubsection{Configuration Module}
\label{Configuration Module}

As mentioned in the formative study, coordinators need to set linguistic target words and configure and adjust children's features (F1, D1). To address this need, we designed the Configuration Module to help the configuration process and enable continuous updates.


First, coordinators set target languages for learning and corresponding vocabulary as target words based on participating children's information, serving as the foundation for generating the collaborative story and ensuring content aligns with learning objectives. Second, coordinators can configure individual features for each child, including Metadata (e.g., cultural background and language proficiency) and Preference (e.g., preferred topic and content). All features are structured as tags, allowing coordinators to flexibly retrieve, add, modify, or delete to improve operational efficiency. Tags are widely used across platforms - social media for identity descriptions and recommendation systems for personalized profiles - enhancing usability and data utility \cite{cai2010personalized, wang2010personalization}. Referring to existing research and practices, our system supports two flexible methods to configure tags:

\begin{itemize}
    \item \textbf{Tag Selection}: The system allows coordinators to configure child features by selecting predefined tags, including Metadata such as "Gender: Female / Male" and Preference such as "Topic Preference: Animals / Sci-Fi", enabling rapid and standardized configuration.
    
    \item \textbf{Tag Input}: To accommodate individualized expression, the system also supports entering tags via text. For example, inputting "Detective Adventure" under "Topic Preference" automatically generates two separate tags, "Detective" and "Adventure".
    
\end{itemize}

To provide a more comprehensive representation of children's features, the system also supports tags for dislikes (e.g., the story topics the child does not like). Combining tag selection with tag input balances standardization and flexibility, establishing a foundation for subsequent individual and common characteristic summarization.

\subsubsection{Individual Characteristic Summarization Module}
\label{Individual Characteristic Summarization Module}

To support personalization, teachers expressed the desire to leverage each child's individual characteristics, particularly for question and material generation (F4, D2). Therefore, we designed the Individual Characteristic Summarization Module to allow coordinators to comprehensively understand each child's preferences beforehand and to support personalized content generation in subsequent modules.


Based on the configuration, this module analyzes and summarizes each child's individual features, presenting concise and clear descriptive sentences to the coordinator via natural language. Expressing descriptions in natural languages serves two purposes: (1) it allows coordinators to easily interpret, review, and edit; and (2) it can be directly used as prompts for LLMs in subsequent content generation. Ultimately, each child will have an Individual Characteristics profile that highlights differences in metadata and preferences. For example, the system might describe Lisa as "highly verbally fluent with a tendency for imaginative twists, occasionally needing guidance to stay focused".



\subsubsection{Common Characteristic Summarization Module}
\label{Common Characteristic Summarization Module}

Teachers also emphasized the importance of identifying common characteristics, which support shared tasks like story framework construction in cross-language collaborative storytelling (F4, D2). To address this, we designed the Common Characteristic Summarization Module, which employs two complementary approaches - semantic matching and semantic reasoning - to extract common characteristics among children, supporting content generation in subsequent shared tasks.


Using LLMs, SparkTales extracts commonalities from configured features through semantic matching and reasoning, guided by a Guideline-driven mechanism \cite{GUIDE} (specifically the Children Characteristic Guidelines) to ensure reliable and interpretable results:



\begin{itemize}
    \item \textbf{Semantic Matching}: There are two approaches: (1) exact matching identifies identical tags among children, for instance, "8 years old" under age or "animals" under topic preference; (2) approximate matching handles similar but non-identical tags, where lexical differences in cross-language contexts make string matching insufficient. Using LLMs, the system employs multilingual concepts and semantics, mapping related tags into unified categories - for example, "superhero" and "kungfu warrior" are recognized as "action-themed heroes".


    \item \textbf{Semantic Reasoning}: For unmatched features, the system leverages children's metadata and a Guideline-driven mechanism \cite{GUIDE} to guide LLMs in inferring commonalities, dynamically adapting guidelines across scenarios to ensure generalizability. For example, by applying exam-level guidelines (e.g., YCT for Youth Chinese Test \cite{yang2024comprehensive} and YLE for Cambridge Young Learners English Tests \cite{YLE}) to infer shared vocabulary and grammar, and preference guidelines (e.g., age and gender) to infer common interests in story topics and plot \cite{collette2019cross, clode2016choose}, the system can identify that two girls of the same age may share similar language use and content preferences.

\end{itemize}

The two approaches complement each other in natural language, first using semantic matching on tags, then metadata-driven reasoning to build holistic commonality profiles - e.g., "Both children are five-year-old boys who play musical instruments (semantic matching), and adventure elements likely appeal to boys this age (semantic reasoning)".


\subsubsection{Collaborative Storytelling Support Module}
\label{Collaborative Storytelling Support Module}
As suggested by D2, teachers expect support for the generation of story framework, questions, and material based on children's individual and common characteristics (F2, F3, D2). Accordingly, we designed the Collaborative Storytelling Support Module in story cloze completion, adaptation, and extension based on children's Individual Characteristics and Common Characteristics, comprising three submodules: Story Framework Generation, Question Generation, and Comprehension-Oriented Material Generation. 


\textbf{Story Framework Generation}: As illustrated in Figure~\ref{fig:monolingual and bilingual collaborative storybooks} (b), cross-language collaborative storytelling typically follows the pattern of "bilingual alternation" and "paragraph cloze". Bilingual alternation, commonly used in bilingual studies, presents target and native languages in alternating paragraphs to facilitate language exchange \cite{2007alternational}, while paragraph cloze, a context-driven approach for vocabulary learning, leaves target words blank to stimulate expression and reinforce vocabulary comprehension \cite{zou2017vocabulary, talebzadeh2012effects}. Therefore, Story Framework Generation generates bilingual story cloze templates based on target words and Common Characteristics.


The generation process follows the segmented progressive story generation method described in \cite{wang2024guiding}, constructing  \textit{premise} and \textit{instruction} variables to guide LLMs with structural control. The \textit{premise} integrates Common Characteristics and target words to outline the story's core topic, while the \textit{instruction} follows the Freytag pyramid structure \cite{ciugerci2024freytag} to cover key narrative stages. Complete prompt details for story generation are provided in Appendix \ref{Appendix_Prompts}. Once the final bilingual story is confirmed, the system removes target words from relevant paragraphs to generate the cloze task.

\textbf{Question Generation}: This module leverages children's Individual Characteristics to dynamically generate personalized guiding questions for three stages, effectively stimulating active thinking and language expression:


\begin{itemize}
    \item \textbf{Question Generation for Story Cloze Completion}: In this stage, questions are generated to guide children in filling blanks with target words, involving vocabulary explanations, synonyms, or related associations (e.g., "Which farm animals can children ride?" for the word "horse").


    \item \textbf{Question Generation for Story Adaptation}: The system generates thought-provoking questions based on the story, encouraging imaginative elaboration (e.g., "If the main character were your favorite Disney princess, what would she do?").


    \item \textbf{Question Generation for Story Extension}: The system uses open-ended questions to extend the story, stimulating imagination and creative expression. It generates different questions for the Storyteller to detail the content (e.g., "What color dress is the princess you mentioned wearing?") and for the Storylistener to assess comprehension (e.g., "Can you retell the story you just heard?")


\end{itemize}

In designing questions, we incorporate Individual Characteristics and prior works on children's storytelling QA methods (FairytaleQA \cite{fantastic} and StorySparkQA \cite{chen2023storysparkqa}) and controllable generation strategies \cite{li2024planning} to guide LLMs. Specifically, following \cite{li2024planning}, we first designed two key variables - \textit{attribute} and \textit{ex\_or\_im} - for the question-generation prompt template (details in Appendix \ref{Appendix_Prompts}):


\begin{itemize}
    \item \textbf{\textit{attribute}}: It refers to dimentional question validated in prior educational research \cite{Assessing}, supporting the multidimensional development of children's narrative comprehension.

    
    \item \textbf{\textit{ex\_or\_im}}: It comprises explicit questions, involving logical reasoning based on existing content with answers directly from the text, and implicit questions, which stimulate imagination and require reasoning for more open-ended answers.


\end{itemize}

The question categories include structured questions based on \textit{attribute} (starting with "who", "where", "what", "when", "why", and "how") and explicit or implicit questions based on \textit{ex\_or\_im}. By combining the values of these two variables in the prompt template, question-generation guidelines can be established for each stage, while the generated questions serve as reference support for coordinators and are not intended to be directly editable. These generated questions are provided as reference support for coordinators and are not intended to be directly editable.



\textbf{Comprehension-Oriented Material Generation}: Based on children's Individual Characteristics, the Comprehension-Oriented Material Generation helps interpret unfamiliar vocabulary or cultural concepts encountered during collaborative storytelling. After inputting keywords, LLMs generate multimodal materials to help coordinators quickly provide explanations to children. 
\textcolor{black}{Coordinators can select culturally aligned explanations based on children's characteristics, allowing unfamiliar, culture-specific concepts to be explained through familiar references. We achieve this by leveraging LLMs' understanding of multiple cultures and designing prompts that specify the child's cultural background (details in Appendix \ref{Appendix_Prompts}). The system then generates analogous explanations based on activities and community practices familiar within the child's own cultural context.
The explanations are presented via multimodal materials, using text for description and images for visualization. For example, when explaining the Chinese square dancing to foreign children, the system can relate it to familiar community dance or group fitness activities and present a cartoon image wherein older adults are dancing together in a park, making the concept more accessible and engaging.}
These system-generated multimodal materials are likewise provided as reference support for coordinators and are not intended to be directly editable.

\subsubsection{Review and Feedback Module}
\label{Review and Feedback Module}
Finally, teachers highlighted the need for feedback to understand children's engagement and refine configurations (F5, D3). In response, we designed the Review and Feedback Module, offering a structured and multidimensional analysis of interaction records for each child, with automatically generated results provided as read-only references, which consists of two components:


\begin{itemize}
    \item \textbf{Engagement Review}: The system summarizes children's participation (e.g., number of questions answered), reflecting the level of language engagement. These data are presented in intuitive formats such as lists and charts, enabling coordinators to assess intuitively and support optimization of their guiding strategies.

    
    \item \textbf{Feature Feedback}: The system dynamically analyzes children's features from their language output, including dynamic Metadata (e.g., language proficiency) and Preference (e.g., preferred topics), and provides descriptive feature explanations, giving coordinators deeper insights into traits and evolving abilities that may be overlooked. For example, repeated mentions of "SpongeBob" are tagged as preference feedback, enabling to suggest of incorporating this character into the next storytelling session.


\end{itemize}


\subsection{Implementation and Example}
\label{SparkTales Implementation}
We implemented SparkTales as a Web application accessible via PCs, tablets, and smartphones, supporting diverse usage scenarios and enhancing flexibility. It integrates Tencent Real-Time Communication (TRTC) services\footnote{https://cloud.tencent.com/document/product/647} for high-quality audio and video interaction and the latest GPT-5\footnote{https://openai.com/gpt-5/} for advanced multimodal generation. 



\begin{figure*}
    \begin{minipage}[t]{0.48\linewidth}
        \centering
        \includegraphics[width=\textwidth]{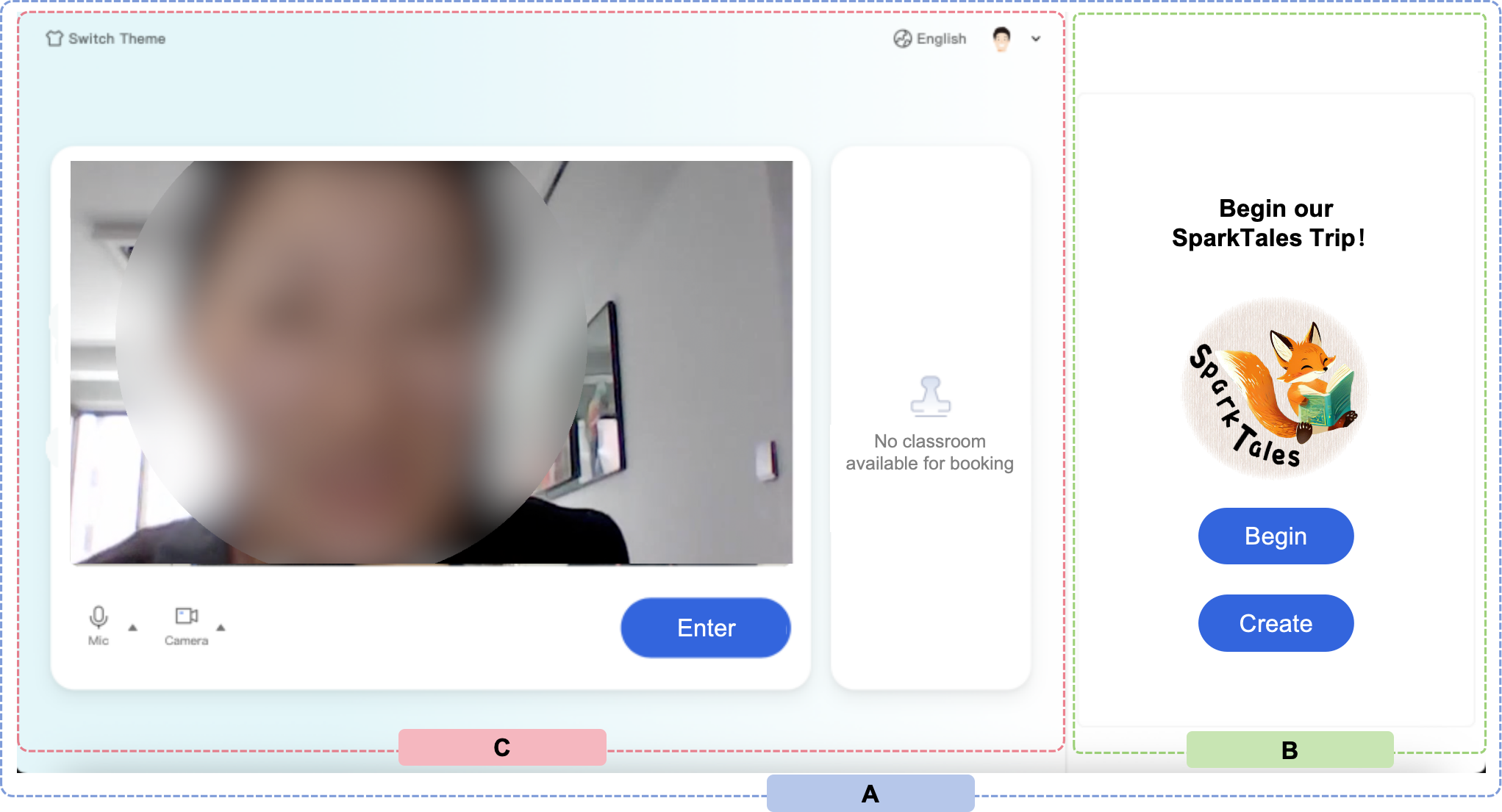}
        \centerline{\small{(a) Preparation interface.}}
    \end{minipage}%
    \hfill
    \begin{minipage}[t]{0.48\linewidth}
        \centering
        \includegraphics[width=\textwidth]{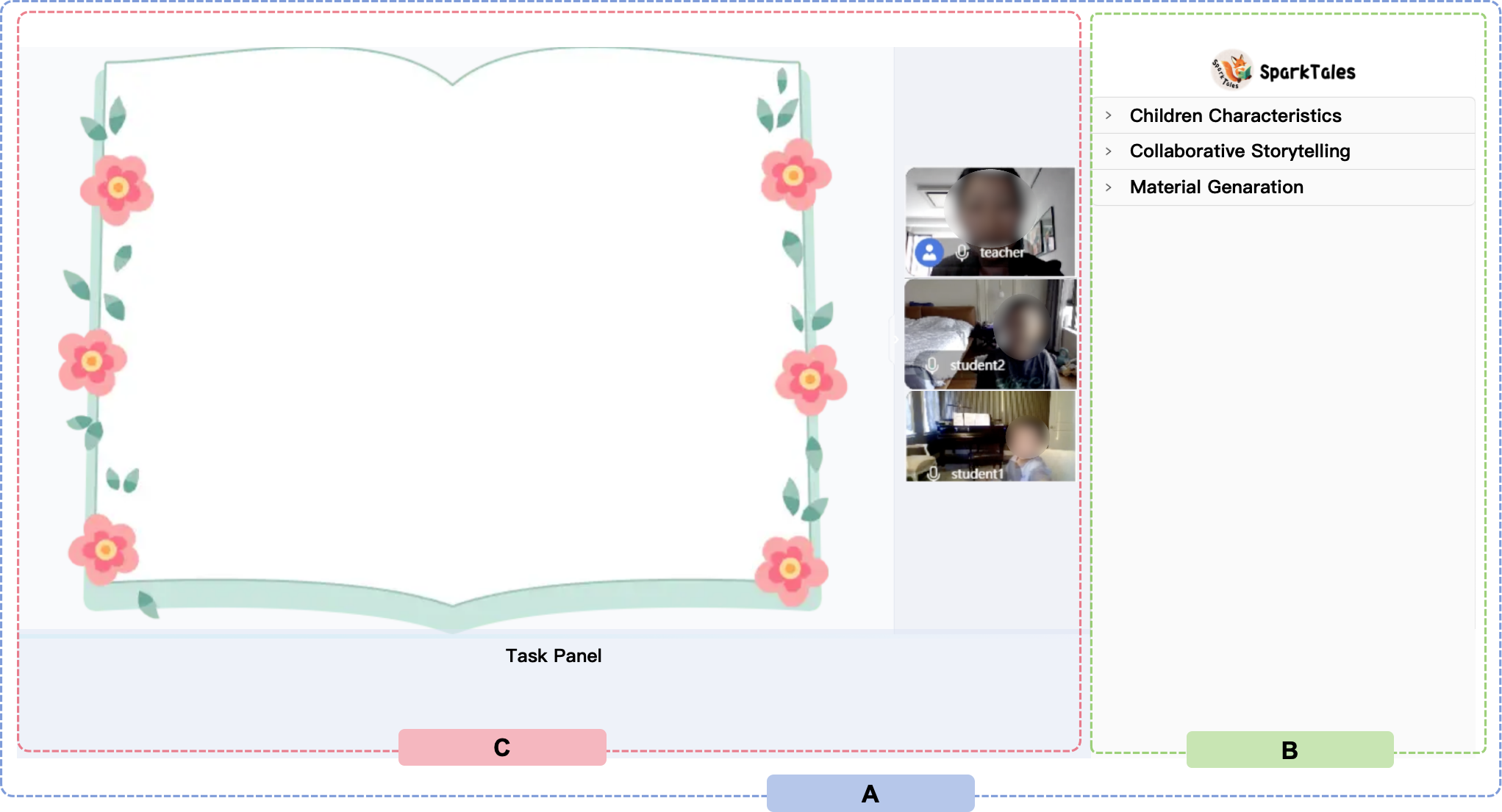}
        \centerline{\small{(b) Storytelling interface.}}
    \end{minipage}
    \vfill
    \caption{SparkTales initiation interface. The area marked with a blue box (A) (abbreviated as \textit{Box A}) represents the coordinator-visible interface; \textit{Box B} indicates the coordinator's configuration panel, visible only to the coordinator; and \textit{Box C} denotes the interface visible to the children.}
    \label{fig: Initiation}
\end{figure*}

To illustrate SparkTales in practice, we present a collaborative storytelling example with one teacher as coordinator and two children, Lisa (learning Chinese) and Lele (learning English). Upon login, the teacher and the children access \textit{Box A} and \textit{Box C} in Figure~\ref{fig: Initiation} (a). After initiating and joining, the teacher and the children are presented with \textit{Box A} and \textit{Box C} in Figure~\ref{fig: Initiation} (b). The whole workflow encompasses three stages: Preparation, Storytelling, and Review.



\subsubsection{Preparation}
\label{Preparation}

\begin{figure*}[t]
    \begin{minipage}[t]{0.26\linewidth}
        \includegraphics[width=\textwidth]{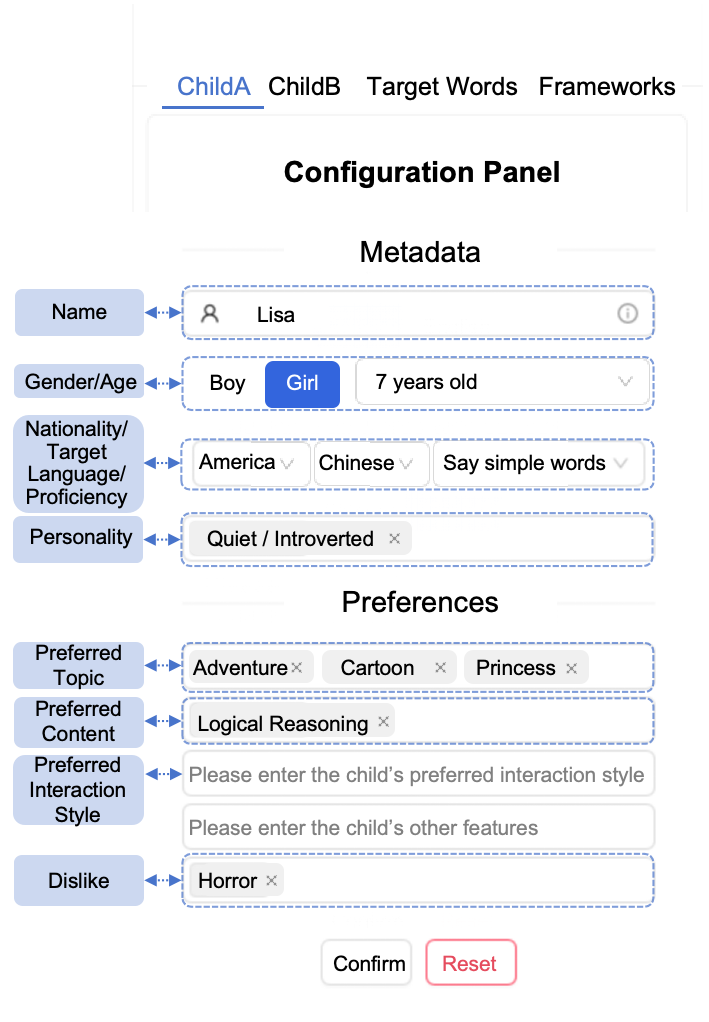}
        \centerline{\small{(a) Child's feature configuration (Lisa).}}
    \end{minipage}%
    \hfill
    \begin{minipage}[t]{0.26\linewidth}
        \includegraphics[width=\textwidth]{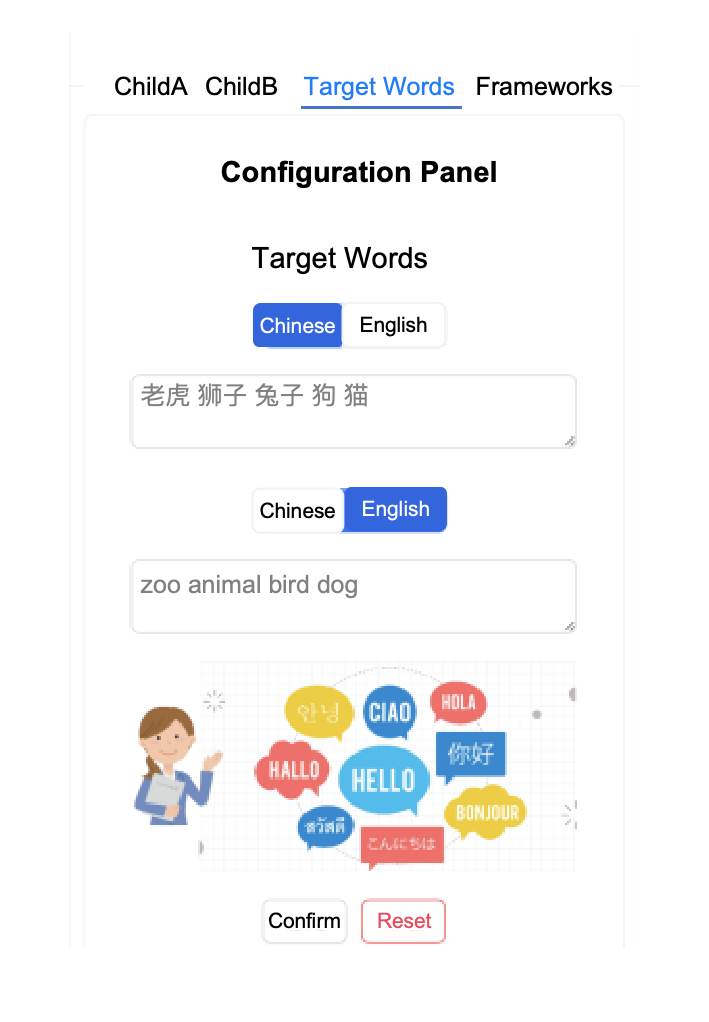}
        \centerline{\small{(b) Target word configuration.}}
    \end{minipage}
        \hfill
    \begin{minipage}[t]{0.26\linewidth}
        \includegraphics[width=\textwidth]{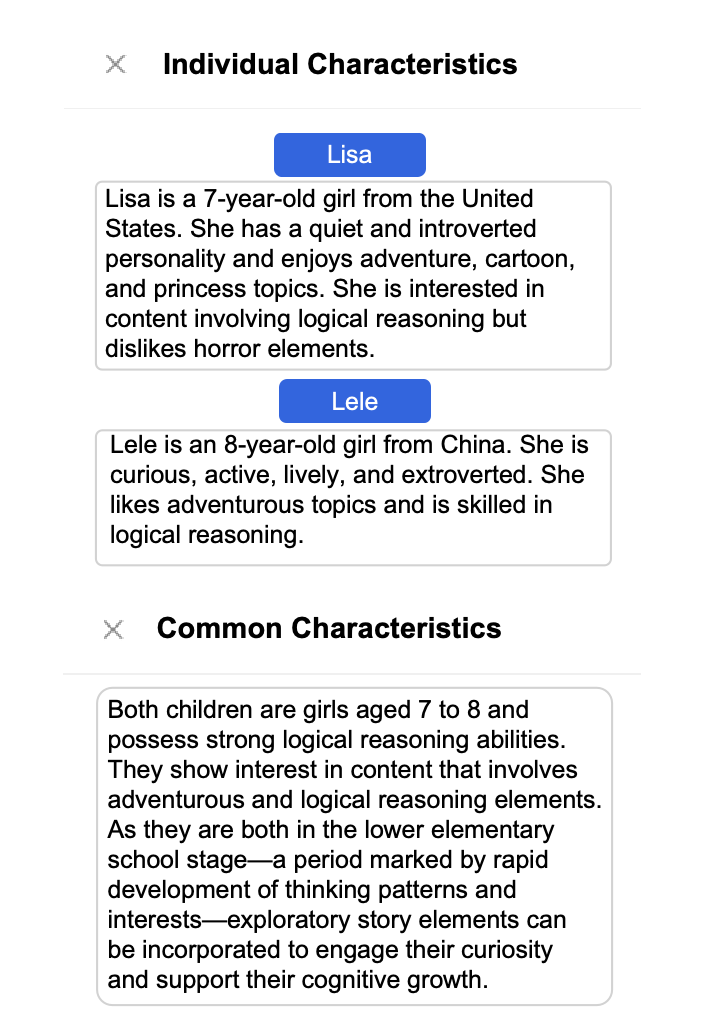}
        \centerline{\small{(c) Characteristic summarization.}}
    \end{minipage}
    \vfill
    \caption{Configuration panel for characteristic configuration and summarization.}
    \label{fig: Configuration}
\end{figure*}

The teacher sets features and target words for Lisa and Lele, as show in Figure~\ref{fig: Configuration} (for brevity, screenshots only display \textit{Box B} in Figure~\ref{fig: Initiation}). 
\textcolor{black}{Figure~\ref{fig: Configuration} (a) shows Lisa's configuration interface, covering Metadata (Gender, Age, Nationality, Language Proficiency, Personality) and Preference (Preferred Topic, Content, Interaction Style).
\begin{CJK}{UTF8}{gbsn}
Figure~\ref{fig: Configuration} (b) presents the bilingual target-word configuration, such as Chinese ("老虎, 狮子, 兔子, 狗, 猫") and English ("zoo, animal, bird, dog").
\end{CJK}}

\subsubsection{Storytelling}
\label{Storytelling}

The system then generates natural-language summaries of individual and common characteristics for Lisa and Lele (Figure~\ref{fig: Configuration} (c)). 
\textcolor{black}{For individual characteristics, the system generates summaries from the configured tags: Lisa (7, U.S.) is quiet, introverted, and prefers adventure, cartoons, princess topics and logical reasoning, while Lele (8, China) is curious, lively, and also enjoys adventure and logical reasoning.
For common characteristics, the system first derives shared traits from the tags, including ages 7–8 and a preference for adventure and reasoning-oriented content. 
Based on the Children Characteristic Guidelines, it then recommends adding exploratory story elements to support their curiosity and cognitive development.}



\begin{figure*}
    \begin{minipage}[t]{0.25\linewidth}
        \centering
        \includegraphics[width=\textwidth]{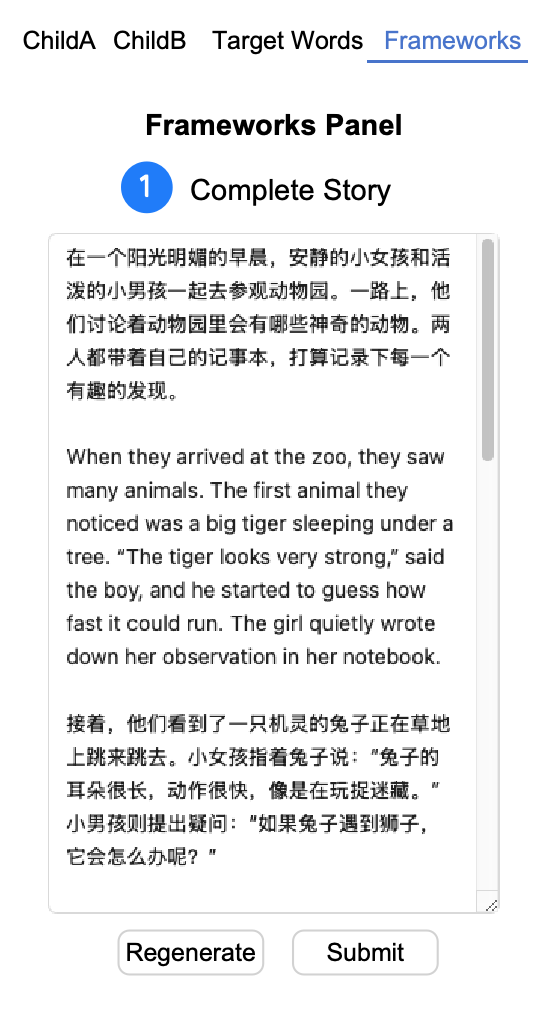}
        \centerline{\small{(a) Complete story.}}
    \end{minipage}%
    \hfill
    \begin{minipage}[t]{0.25\linewidth}
        \centering
        \includegraphics[width=\textwidth]{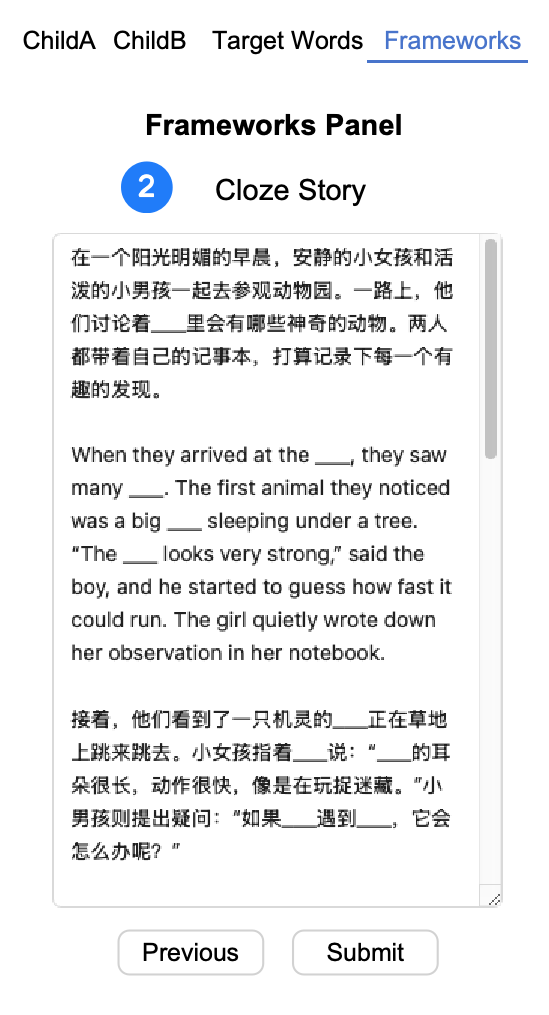}
        \centerline{\small{(b) Cloze story.}}
    \end{minipage}
    \hfill
    \begin{minipage}[t]{0.27\linewidth}
        \centering
        \includegraphics[width=\textwidth]{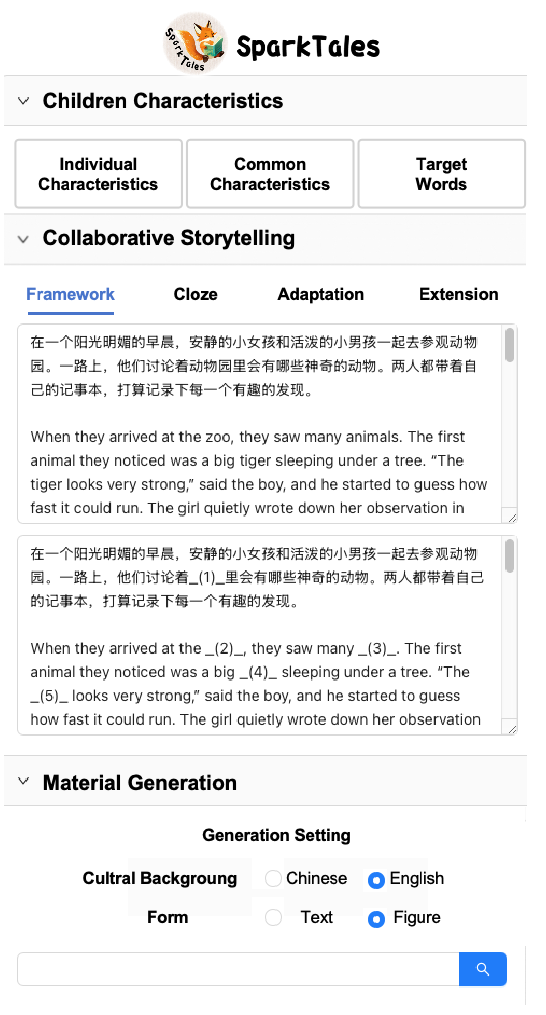}
        \centerline{\small{(c) Finalized cloze story.}}
    \end{minipage}
    \vfill
    \caption{Configuration panel for story framework generation.}
    \label{fig: Story framework generation}
\end{figure*}

\textbf{Story Framework Generation}: 
\textcolor{black}{With target words related to "zoo" and the children's shared interests in adventure and logical reasoning, the system generates a bilingual story centered on a "zoo" topic, depicting a boy and a girl exploring the zoo on an adventure (Figure~\ref{fig: Story framework generation} (a)).}
The teacher could edit the text or click "Regenerate" for a new version; once confirmed, clicking "Submit" converts the story into a cloze version (Figure~\ref{fig: Story framework generation} (b)), which remains editable before final submission. Figure~\ref{fig: Story framework generation} (c) shows the finalized cloze story in the coordinators' configuration panel.

\begin{figure*}[t]
    \begin{minipage}[t]{0.25\linewidth}
        \centering
        \includegraphics[width=\textwidth]{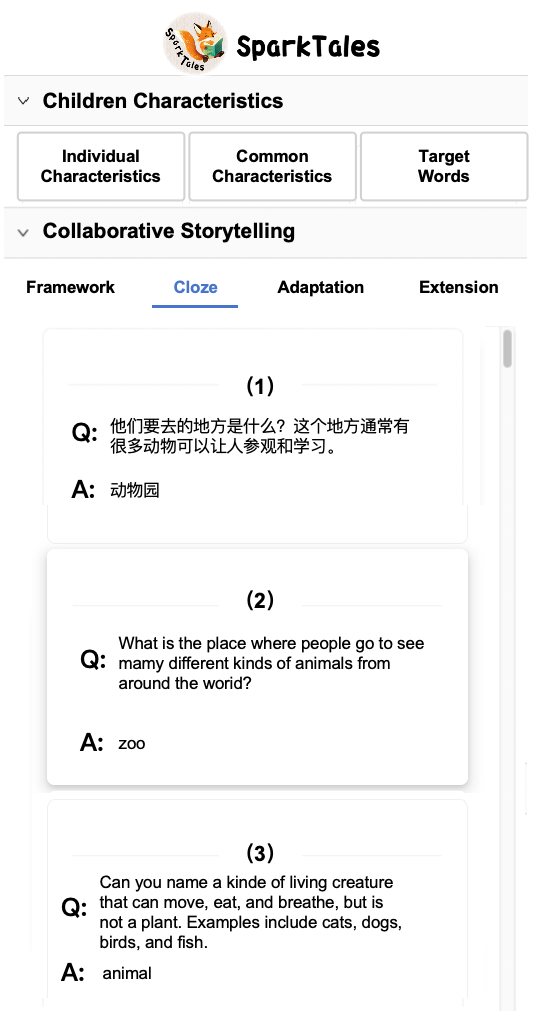}
        \centerline{\small{(a) Story cloze completion phase.}}
    \end{minipage}%
    \hfill
    \begin{minipage}[t]{0.25\linewidth}
        \centering
        \includegraphics[width=\textwidth]{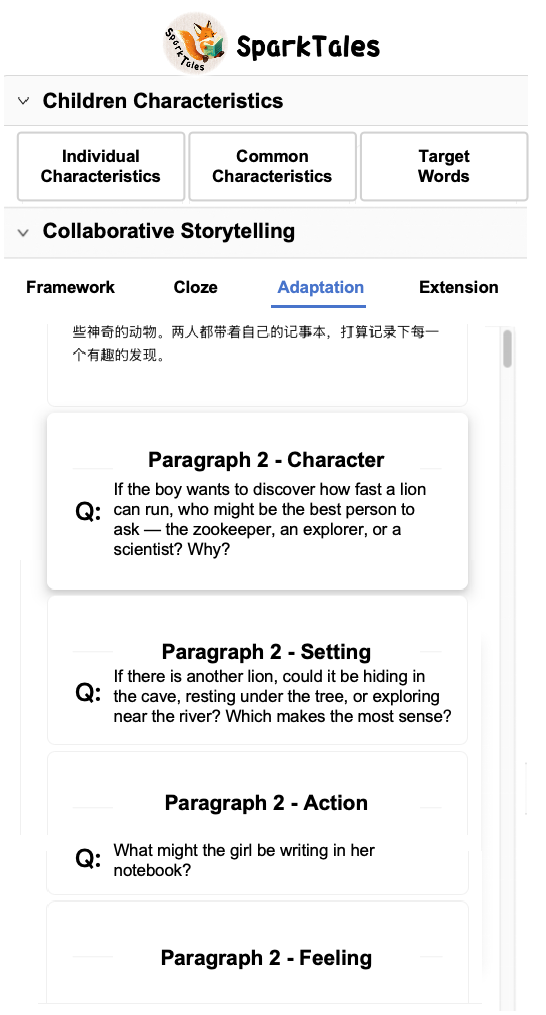}
        \centerline{\small{(b) Story adaptation phase.}}
    \end{minipage}%
    \hfill
    \begin{minipage}[t]{0.25\linewidth}
        \centering
        \includegraphics[width=\textwidth]{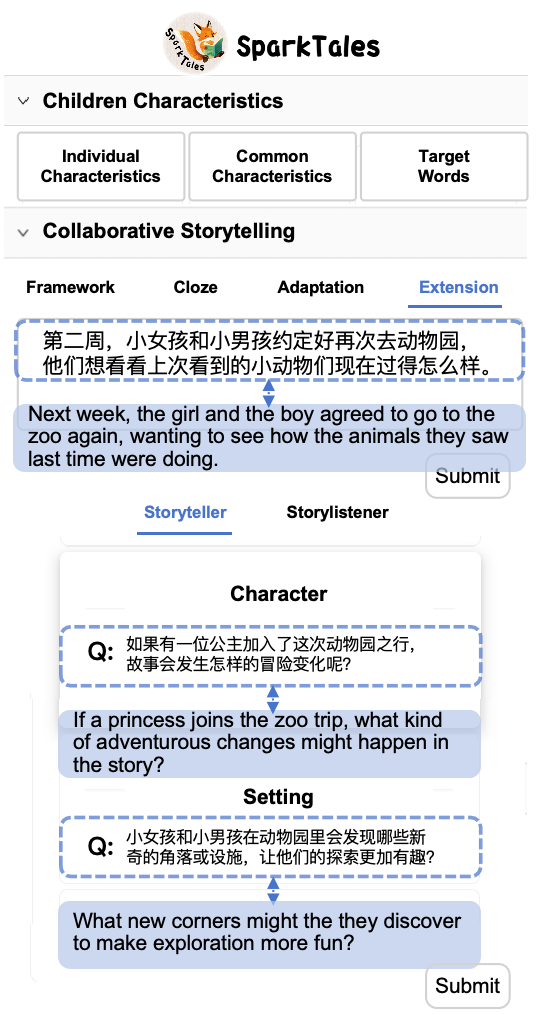}
        \centerline{\small{(c) Story extension phase.}}
    \end{minipage}
    \vfill
    \caption{Configuration panel for question generation.}
    \label{fig: Question generation}
\end{figure*}

\begin{figure*}
  \centering
  \includegraphics[width=0.8\linewidth]{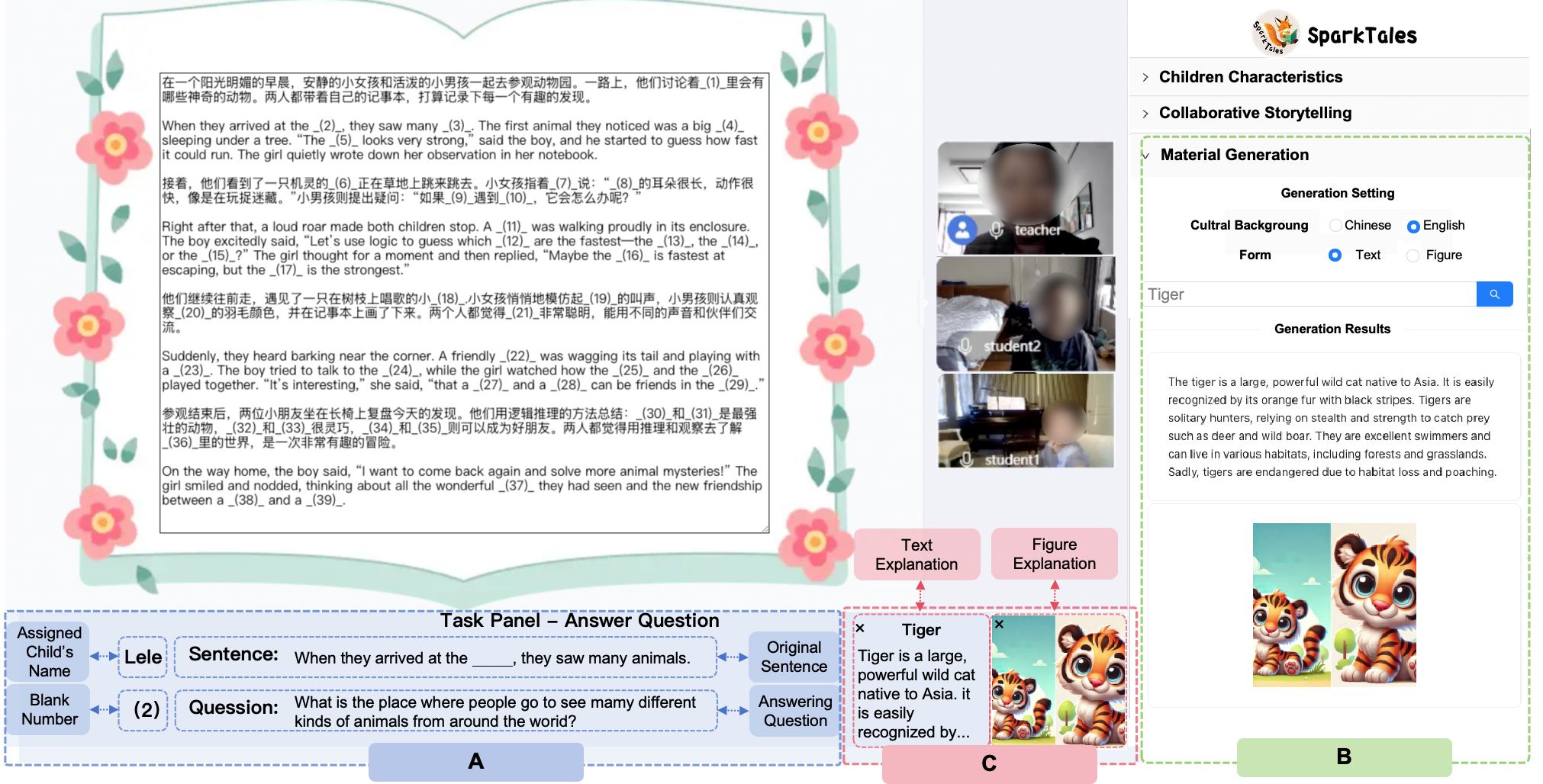}
  \caption{SparkTales interface during storytelling. \textit{Box A} represents the shared task panel for children, visible to both children and the coordinator; \textit{Box B} indicates the coordinator's configuration panel, visible only to the coordinator; and \textit{Box C} denotes the material generation display, visible to both children and the coordinator.}

  \label{fig: material generation}
\end{figure*}

\textbf{Story Cloze Completion}: During cloze completion, the teacher guides the children to take turns filling in the blanks. If unsure how to frame a question, the teacher could refer to system-generated questions (Figure~\ref{fig: Question generation} (a)) corresponding to each blank. 
\textcolor{black}{For example, for blank (2) corresponding to the word "zoo", the teacher selects the question "What is the place where people go to see many different kinds of animals from around the world?"} and presents it to the children (showing in shared task panel, \textit{Box A} in Figure~\ref{fig: material generation}). 
The system also displays the assigned child's name (Lele) and the blank number (2). The teacher could insert the answer into the text, progressively enriching the framework.

\textbf{Story Adaptation}: In the adaptation stage, the teacher could also seek support from SparkTales when needed, as it generates questions across seven dimensions for each paragraph (Figure~\ref{fig: Question generation} (b)). For example, the teacher could select and present a "Character" question for Lele, specifically targeting the second paragraph, which aligns with Lele's logical reasoning preference, \textcolor{black}{such as: "If the boy wants to discover how fast a lion can run, who might be the best person to ask — the zookeeper, an explorer, or a scientist? Why?"} Then, the teacher could revise the text based on Lele's responses.


\textbf{Story Extension}: During story extension, Lisa and Lele alternate as Storyteller and Storylistener. When Lisa is the Storyteller, the system transcribes her narration — \textcolor{black}{"Next week, the girl and the boy agreed to go to the zoo again, wanting to see how the animals they saw last time were doing."} — and provides the teacher with seven-dimensional questions tailored to her characteristics for elaboration (Figure~\ref{fig: Question generation} (c)), which are then synchronized to Lisa after selection. \textcolor{black}{For example, one question could be: "What new corners might they discover to make the exploration more fun?"}
Meanwhile, Lele, as Storylistener, understands Lisa's expression and answers comprehension questions.

Additionally, the material generation panel (\textit{Box B} in Figure~\ref{fig: material generation}) produces multimodal resources to support comprehension throughout the activity. For instance, the teacher can input "tiger" to generate images and text, and present them to the children (\textit{Box C} in Figure~\ref{fig: material generation}).


\subsubsection{Review}
\label{Review}

\begin{figure*}
  \centering
  \includegraphics[width=0.55\linewidth]{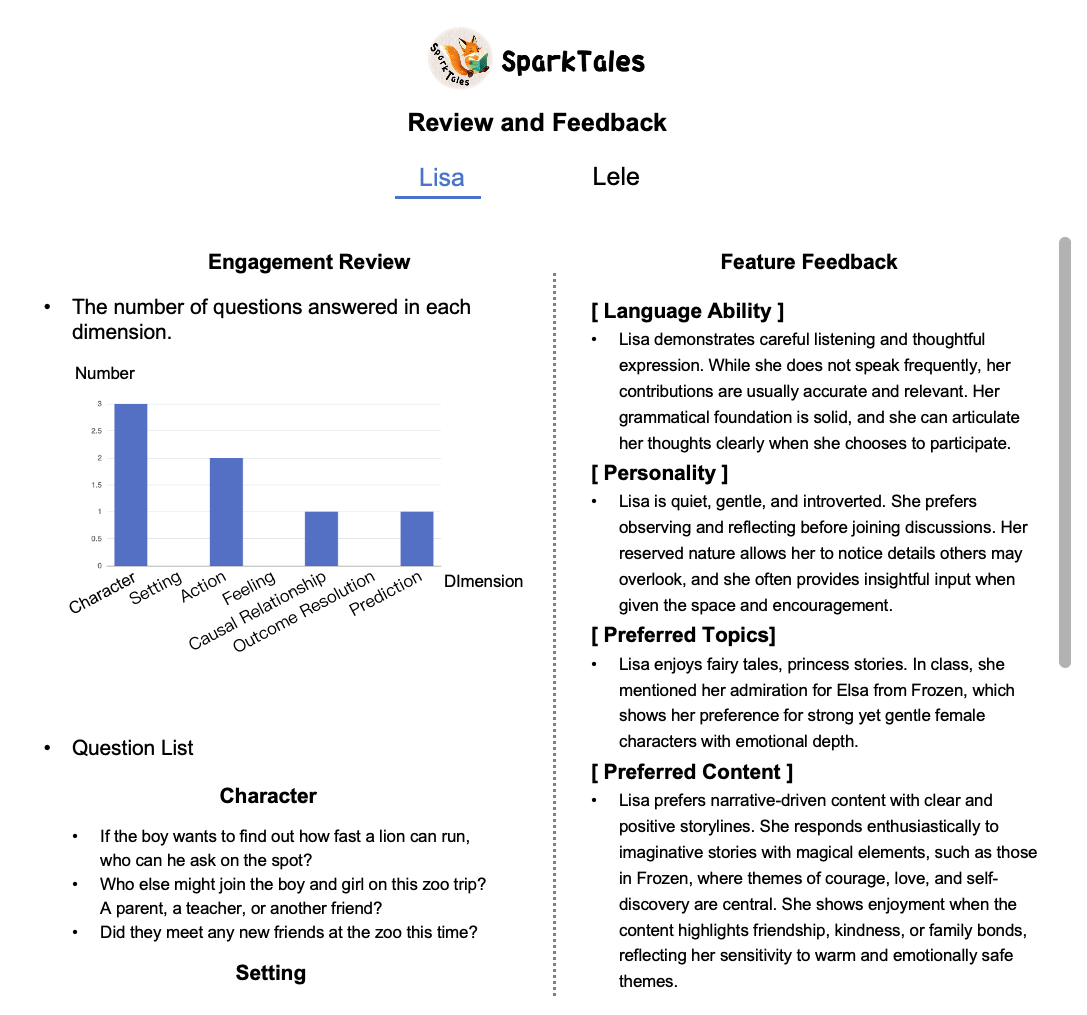}
  \caption{Review and feedback interface.}
  \label{fig:review}
\end{figure*}

Figure~\ref{fig:review} shows Lisa's engagement review and feature feedback after the activity. 
\textcolor{black}{Engagement Review presents statistics of the questions Lisa answered, showing frequent response to "Character" questions and none for "Setting", "Feeling", or "Outcome Resolution". The system also lists the specific questions she answered. 
Feature Feedback summarizes her preferences, highlighting repeated mentions of Elsa from Frozen, indicating a strong interest in related characters and themes.}



\subsection{Summary of Characteristics}
\label{Summary of Characteristic}
To summarize, the characteristics of SparkTales are as follows:

\begin{itemize}
    \item \textcolor{black}{\textbf{Integrate Individual and Common Characteristics (D1, D2)}: SparkTales leverages both the Individual Characteristic Summarization Module and the Common Characteristic Summarization Module to inform story framework generation, cloze completion, adaptation, and extension. By integrating individual traits to sustain each child's curiosity and common traits to enhance group resonance, SparkTales can more effectively support children's active participation.}
    
    
    \item \textcolor{black}{\textbf{Capture and Update Children's Features (D1, D3, D4)}: SparkTales enables coordinators to set children's features before collaboration through the Configuration Module and dynamically update them based on the Review and Feedback Module, helping coordinators track and analyze children's changes.}
    
    
    \item \textcolor{black}{\textbf{Balance AI Support with Coordinator Control (D4)}: SparkTales uses the Collaborative Storytelling Support Module to offer real-time assistance while preserving coordinators' control over operations and content adjustments. This balance of AI support and human coordination ensures appropriate AI involvement and mitigates potential risks to children.}
    
    
    \item \textcolor{black}{\textbf{Modular Design}: SparkTales adopts a modular architecture that decouples its five modules, enabling independent updates and ensuring flexibility to adapt to evolving requirements and advancements.}
    
\end{itemize}

\section{Evaluation}
\label{Evaluation}
Regarding the design goals, we conducted extensive evaluations of SparkTales to first examine whether the system can effectively support coordinators in cross-language collaborative storytelling, and then to explore how such support can contribute to improving children's engagement. We address the following two research questions:

\begin{itemize}
    \item RQ1: How can SparkTales help coordinators conduct cross-language collaborative storytelling? 
    \item RQ2: How can SparkTales help improve children’s engagement?
\end{itemize}

\subsection{Settings}
\label{Settings}

\subsubsection{Evaluation Metrics}
\label{Evaluation Metrics}
\textcolor{black}{To address these research questions, we employed quantitative and qualitative evaluations based on the Technology Acceptance Model (TAM) \cite{marangunic2015technology} and Verbal Engagement Evaluation \cite{xu2021same, vukelich1976development, westerveld2017oral}, focusing on both coordinators' using experiences and children's language participation.}


\textcolor{black}{\textbf{Evaluation Metrics for RQ1}:
To address RQ1, we evaluated coordinators' acceptance of the system with reference to TAM \cite{marangunic2015technology}, grounded in psychological and planned behavior theories. The evaluation encompassed three core dimensions: \textbf{Function}, which examines whether the system provides coordinators with complete capabilities for configuration, generation, feedback, and process control; \textbf{Performance}, which assesses the coordinators' perceived accuracy and appropriateness of system outputs as well as their acceptance; and \textbf{Usability}, which focuses on ease of use, accessibility, clarity of interface and content presentation, and the sense of control during use. 
}

\textbf{\textcolor{black}{Evaluation Metrics for RQ2}}: We also evaluated the impact of SparkTales on children's verbal engagement during collaborative storytelling. Drawing on relevant literature in children's storytelling \cite{xu2021same, vukelich1976development, westerveld2017oral}, we analyzed verbal engagement through the following six dimensions.

\begin{itemize}
    \item \textbf{Number of Questions Answered}: It represents how many times a child responded to the coordinator's questions during the interaction.

    \item \textbf{Productivity}: It is measured by the total number of words produced in a child's responses. All words in each answer are counted, excluding meaningless expressions like "uh", "um", "aha", are not included.

    \item \textbf{Lexical Diversity}: It is assessed by the number of unique words appearing in a child's responses. Repeated words are excluded, and only distinct words are counted.

    \item \textbf{Topical Relevance}: It is coded on a three-level scale, measuring whether each answer follows the conversation context. A direct response to the question will receive a score of 2, an indirect but topically related response will receive 1, and an irrelevant response to the question or topic will receive 0.

    \item \textbf{Intelligibility}: It is rated on a 0 to 2 scale, following \cite{flipsenjr2002longitudinal}. If the answer is clear and fully understandable, it will receive 2. If most of the answer is intelligible except for one or two words, it will receive 1. And if the answer is mostly unintelligible with less than 50\% of the content understood, it will receive a score of 0. 

    \item \textbf{Accuracy}: It is coded as a binary variable indicating correct (1) or incorrect (0).
\end{itemize}

\subsubsection{Participants}
\label{eva: Participants}

\begin{table*}[ht] 
\caption{Demographics of evaluation participants (teachers).}
\label{tab:Summary of evaluation participants for teachers}
\centering
\resizebox{\linewidth}{!}{
\begin{tabular}{c|c|ccccccccc}
\midrule
& \textbf{\begin{tabular}[c]{@{}c@{}}Teacher \\ ID\end{tabular}}  & \textbf{Age} & \textbf{Gender} & \textbf{Education} & \textbf{Major} & \textbf{\begin{tabular}[c]{@{}c@{}}Age Group \\ Taught\end{tabular}} & \textbf{\begin{tabular}[c]{@{}c@{}}Language \\ Taught\end{tabular}}& \textbf{\begin{tabular}[c]{@{}c@{}}Years of \\ Online Teaching\end{tabular}}&
\textbf{\begin{tabular}[c]{@{}c@{}}Frequency of  \\Collaborative Storytelling\end{tabular}} & \textbf{\begin{tabular}[c]{@{}c@{}}Whether Participated\\ in Formative Study\end{tabular}} \\ \midrule
\textbf{G1}  & T1         & 18-24 & Male    &  Master's Degree      &  TCSOL         & \begin{tabular}[c]{@{}c@{}}High School, \\ College\end{tabular}      &     English, Chinese         & 1-2 years         &  About once a month                        &  N\\ \midrule
\textbf{G2}  & T2   &      35-44                                        & Female            & Master's Degree                              & English Education    & \begin{tabular}[c]{@{}c@{}}Preschool, \\Primary School\end{tabular}   &  English, Chinese & 6–10 years & 1–2 times per week     &              Y                                                                     \\ \midrule
\textbf{G3}  & T3         &25-34                                        & Female            & Master's Degree                              & Computer Science    & All Age Groups   &  English, Chinese, French & 6–10 years & 1–2 times per week        &   Y                                                                                \\ \midrule
\textbf{G4}  & T4         &35-44                                        & Female            & Master's Degree                              & TCSOL    & \begin{tabular}[c]{@{}c@{}}Primary School, \\Middle School\end{tabular}   &  English, Chinese & 6–10 years & 3 or more times per week        &   N                                                                                   \\ \midrule
\textbf{G5}  & T5         &25-34                                        & Female            & Master's Degree                              & TCSOL    & Primary School   &  English, Chinese & 3-5 years & About once a month        &   N\\ \midrule
\textbf{G6}  & T6         &35-44                                        & Female            & Master's Degree                              & TCSOL    & \begin{tabular}[c]{@{}c@{}}Primary School,\\ Middle School\end{tabular}   &  English, Chinese & More than 10 years  & 1–2 times per week        &   Y\\ \midrule
\textbf{G7}  & T7         &18-24                                        & Female            & Master's Degree                              & TCSOL    & All Age Groups   &  English, Chinese & 1-2 years & About once a month        &   N\\ \midrule
\textbf{G8}  & T8         &25-34                                        & Female            & Master's Degree                              & TCSOL    & All Age Groups   &  English, Chinese & 3-5 years & 3 or more times per week    &                     Y                                                              \\ \midrule
\end{tabular}
}
\raggedright
\footnotesize
Note: TCSOL = Teaching Chinese to Speakers of Other Languages.

\end{table*}

\begin{table*}[ht] 
\caption{Demographics of evaluation participants (children).}
\label{tab:Summary of evaluation participants for children}
\centering
\resizebox{\linewidth}{!}{
\begin{tabular}{ccccccccccccc}
\midrule
\multicolumn{1}{c|}{\multirow{2}{*}{\textbf{}}} & \multicolumn{6}{c|}{\textbf{Child A}} & \multicolumn{6}{c}{\textbf{Child B}} \\ 
\midrule 

\multicolumn{1}{c|}{} & \multicolumn{1}{c|}{\textbf{Child ID}} & \textbf{Age} & \textbf{Gender} & \textbf{Country} & \textbf{Learning Language} & \multicolumn{1}{c|}{\textbf{Learning Proficiency}} & 
\multicolumn{1}{c|}{\textbf{Child ID}} & \textbf{Age} & \textbf{Gender} & \textbf{Country} & \textbf{Learning Language} & \textbf{Learning Proficiency} \\ \midrule

\multicolumn{1}{c|}{\textbf{G1}} & \multicolumn{1}{c|}{C1-1} & 11 & Girl & China &English & \multicolumn{1}{c|}{A2} & \multicolumn{1}{c|}{C1-2} & 11 & Girl & Canada & Chinese & A2  \\ \midrule

\multicolumn{1}{c|}{\textbf{G2}} & \multicolumn{1}{c|}{C2-1} & 8 & Girl & China &English & \multicolumn{1}{c|}{A1} & \multicolumn{1}{c|}{C2-2} & 10 & Girl & Canada & Chinese & A2  \\ \midrule

\multicolumn{1}{c|}{\textbf{G3}} & \multicolumn{1}{c|}{C3-1} & 10 & Boy & China &English & \multicolumn{1}{c|}{B1} & \multicolumn{1}{c|}{C3-2} & 9 & Boy & America & Chinese & A2  \\ \midrule

\multicolumn{1}{c|}{\textbf{G4}} & \multicolumn{1}{c|}{C4-1} & 10 & Girl & China &English & \multicolumn{1}{c|}{A2} & \multicolumn{1}{c|}{C4-2} & 10 & Girl & America & Chinese & B1  \\ \midrule

\multicolumn{1}{c|}{\textbf{G5}} & \multicolumn{1}{c|}{C5-1} & 8 & Boy & China &English & \multicolumn{1}{c|}{A1} & \multicolumn{1}{c|}{C5-2} & 7 & Boy & America & Chinese & B1  \\ \midrule

\multicolumn{1}{c|}{\textbf{G6}} & \multicolumn{1}{c|}{C6-1} & 8 & Boy & China &English & \multicolumn{1}{c|}{A2} & \multicolumn{1}{c|}{C6-2} & 7 & Boy & America & Chinese & B1  \\ \midrule

\multicolumn{1}{c|}{\textbf{G7}} & \multicolumn{1}{c|}{C7-1} & 9 & Girl & China &English & \multicolumn{1}{c|}{A2} & \multicolumn{1}{c|}{C7-2} & 8 & Girl & America & Chinese & A1  \\ \midrule

\multicolumn{1}{c|}{\textbf{G8}} & \multicolumn{1}{c|}{C8-1} & 7 & Girl & China &English & \multicolumn{1}{c|}{A1} & \multicolumn{1}{c|}{C8-2} & 7 & Girl & America & Chinese & A1  \\ \midrule
\end{tabular}
}
\end{table*}

We recruited 8 groups for evaluation, each consisting of two children and one teacher acting as the coordinator. A total of 8 teachers participated, including 4 from the formative study and 4 newly recruited following the same screening criteria as Section~\ref{formative_Participants}, to enhance the reliability and generalizability. Teachers were briefed on SparkTales' design objectives and evaluation procedures, provided informed consent, and received 100 yuan as compensation. For children, parents or guardians provided consent after being informed of the study's purpose and procedures. Finally, we recruited 16 children aged 7 to 11 from diverse backgrounds, primarily from China, the United States, and Canada, with English and Chinese as their main languages of learning. Child pairings were determined based on the willingness of both children and their guardians, considering factors such as age and gender, while referring to commonly used child-pairing principles in cross-language collaborative storytelling aimed at promoting positive interactions \cite{howe2012children, leaper1991influence, belly2024enhancing}. Detailed information about the participants is provided in Table~\ref{tab:Summary of evaluation participants for teachers} and Table~\ref{tab:Summary of evaluation participants for children}, respectively (where G indicates groups, T indicates teachers, and C indicates children). Since the recruited children did not undergo formal language tests such as YCT or YLE, we combined the information provided by their parents and teachers assessments to map their proficiency onto the globally recognized Common European Framework of Reference for Languages (CEFR) scale \cite{council2001common} (levels A1, A2, B1, B2, C1, C2, with C2 representing the highest proficiency), defining the Language Proficiency features.
 
\subsubsection{Procedure}
\label{eva: Procedure}
The evaluation comprised three stages: (1) pre-activity, teachers received brief training and input children's basic information into the system (e.g., age, language proficiency, and interests, optionally provided by their parents), taking approximately 30 minutes; (2) activity, each teacher guided a group of children through a full collaborative storytelling task, including story framework generation, story cloze completion, story adaptation, and story extension. The procedure lasted about 30 minutes, but was adjustable as needed; (3) post-activity, teachers completed a 5-point Likert questionnaire to evaluate the system, 
and participated in 30-minute semi-structured interviews to supplement and enrich the quantitative results. The entire process was audio-recorded and analyzed by researchers, following the methodology as Section~\ref{formative_procedure}. Researchers interacted with teachers only before and after the activities, without intervening in teacher-child interactions, and all procedures followed ethical guidelines to ensure privacy and safety (see Section~\ref{Ethical Consideration}).

To evaluate the impact on children's engagement, we coded their verbal responses to the coordinators' questions during the activities based on six metrics of verbal engagement outlined in Section~\ref{Evaluation Metrics}. Two researchers independently coded the data, compared results and resolved discrepancies through discussion. The inter-class correlation coefficients (ICCs) were computed to ensure reliability: number of questions answered (1), productivity (1), lexical diversity (1), topical relevance (0.91), accuracy (1), and intelligibility (0.87).

\subsection{Results}
\label{Results}

\subsubsection{\textcolor{black}{RQ1: How can SparkTales help coordinators conduct cross-language collaborative storytelling?}}

\textcolor{black}{To investigate coordinators' perception of SparkTales in supporting cross-language collaborative storytelling, we evaluated the system and its core modules through questionnaires and interviews, focusing on three dimensions: \textbf{Function}, \textbf{Performance}, and \textbf{Usability}.
The five-point Likert scale questionnaire demonstrated high internal consistency, yielding a Cronbach's alpha of $\alpha = 0.949$ \cite{cho2015cronbach}, which exceeds the acceptable threshold of 0.7 \cite{taber2018use}.}

\textcolor{black}{\noindent\textbf{Overall Results:}
 regarding \textbf{Function}, the average score was 4.53, indicating that teachers considered the system's functionalities comprehensive for supporting collaborative storytelling.
Participants confirmed that the configuration, generation, feedback, and process control features met their needs. For instance, teachers noted that "\textit{inputting learner profiles is helpful}" (T8), "\textit{the generated questions cover multiple dimensions}" (T1), and "\textit{the system provides extensive functions while preserving teacher control}" (T8).
Regarding \textbf{Performance}, the average score reached 4.47, reflecting consensus that the generated content was accurate and appropriate.
Teachers highlighted the system's ability to personalize content, remarking that "\textit{the summarization feels quite accurate}" (T5) and "\textit{one child's interest in the stone exhibition led to the generation of appropriate related questions}" (T6).
For \textbf{Usability}, the average score was 4.39, with teachers consistently describing SparkTales as user-friendly and intuitive. They emphasized that the configuration process was "\textit{easy to use with tags and keywords}" (T3), story generation "\textit{operated smoothly}" (T3), question and material generation were "\textit{highly flexible}" (T8), and the feedback panel was "\textit{clear and legible}" (T1).
Given SparkTales' strengths in functional completeness, performance, and usability, teachers reported that the system effectively supports cross-language storytelling and reduces workload. They emphasized that it can "\textit{greatly save review time}" (T7), "\textit{alleviate organizational stress}" (T5), and ultimately "\textit{reduce my workload}" (T3).}

\textcolor{black}{\noindent\textbf{Key Module and Mechanism Analysis:}
SparkTales is designed to support coordinators in three critical tasks during cross-language collaborative storytelling \textbf{(D2)}: \textbf{Story Framework Generation}, \textbf{Question Generation}, and \textbf{Material Generation}.
We conducted an in-depth evaluation of each task.
For \textbf{Story Framework Generation} (Avg: 4.21; Function: 4.38; Performance: 4.31; Usability: 3.94), teachers agreed that the frameworks were contextually appropriate and well-structured. As one teacher noted, "\textit{the story framework aligns very well with my needs and the children's interests}" (T2).
For \textbf{Question Generation} (Avg: 4.27; Function: 4.50; Performance: 4.25; Usability: 4.06), teachers found the output rich and comprehensive, noting that the system "\textit{provides a wide range of questions to help advance the activity}" (T5).
For \textbf{Material Generation} (Avg: 4.35; Function: 4.25; Performance: 4.38; Usability: 4.44), teachers reported that this module directly improved activity fluency due to its timeliness and efficiency. One teacher remarked, "\textit{the speed is quite fast; I can get exactly what I need without the trouble of manual searching}" (T1).
Based on interview feedback, we attribute these high ratings to two primary factors.
First, the generated content aligns closely with activity goals, such as target vocabulary and storylines. This is enabled by the strong cross-cultural comprehension and reasoning capabilities of LLMs, which support effective plot construction, question design, and explanation generation.
Second, the dynamic integration of individual and common characteristics allows SparkTales to balance personalization with shared context. For tasks involving individual differences, the system offers personalization based on learner profiles, such as generating questions tailored to specific language proficiencies and interests. Conversely, for shared tasks—such as understanding the story framework or mastering target words—the system provides generic guidance based on common characteristics. Specifically, SparkTales leverages this integration as follows:}

\begin{itemize}
    \item \textcolor{black}{\textbf{High Accuracy:} Across the three tasks, teachers consistently reported that the generated content was highly aligned with children's language levels and interest profiles, demonstrating SparkTales' precision in extracting and summarizing both individual and common characteristics. Teachers rated the accuracy of the Individual Characteristic Summarization Module and Common Characteristic Summarization Module with average scores of 4.63 and 4.88, respectively. These ratings indicate that the generated content effectively addresses both personalized needs and shared attributes. As T7 explained, "\textit{whether personalized or common, the summarized content fully aligns with the configured features}," and T1 observed, "\textit{the generated content requires almost no manual adjustment, allowing children to respond smoothly during interactions}."}

    \item \textcolor{black}{\textbf{Balancing Personalization and Commonality:} Teachers reported that the system effectively balances individual and common characteristics, enabling generated content—such as story frameworks—to seamlessly integrate personal traits with shared attributes.
    This balance allows the storytelling process to cater to each child's specific language proficiency and interests while simultaneously incorporating the pair's shared features, thereby sustaining smooth and cohesive interaction.
    As one teacher noted, "\textit{the system shows me the children's common interests while also highlighting each child's individuality, making the story interaction much smoother}" (T3).
    These insights highlight that achieving a balance between personalization and commonality is a core strength that enables teachers to effectively facilitate collaborative storytelling.}

    \item \textcolor{black}{\textbf{High Functional Completeness and Usability:} Teachers also reported that SparkTales demonstrates high functional completeness and usability, allowing them to configure individual and common characteristics to support the three tasks effectively. They rated the functional completeness of the Individual and Common Characteristic Summarization Modules with average scores of 4.62 and 4.75, respectively. Teachers noted that "\textit{the story framework, based on common characteristics, resonates with children and facilitates collaborative task completion}" (T4); "\textit{each question aligns with the child's preferences and characteristics}" (T7); and "\textit{the explanatory materials fit individual preferences, making it easier to organize activities}" (T6).
    Furthermore, both modules received high usability ratings of 4.88, reflecting a clear, structured presentation that enables teachers to quickly understand and adjust content. As T2 noted, "\textit{I can easily view children's individual and common features}," and T5 added, "\textit{the natural language descriptions are clear and easy to understand}."}

\end{itemize}

\textcolor{black}{Furthermore, we conducted additional analyses to ensure our evaluations addressed the design goals outlined in Section \ref{Design Goals}. Teachers' overall ratings for the Configuration Module and Review and Feedback Module were 4.53 (Function: 4.63; Performance: 4.75; Usability: 4.50) and 4.61 (Function: 4.56; Performance: 4.56; Usability: 4.56), respectively. Participants generally agreed that these modules effectively support the configuration of learner profiles \textbf{(D1)} and the generation of post-session review and feedback \textbf{(D3)}, thereby enhancing the efficiency of preparation and reflection.
Meanwhile, the overall controllability of the system received an average score of 4.47 (Function: 4.50; Performance: 4.63; Usability: 4.38), with teachers noting that the system offered sufficient flexibility and user control \textbf{(D4)}.
}

\textcolor{black}{\textbf{Limitations \& Future Improvements:} Although SparkTales performed well in terms of functionality, performance, and usability—significantly reducing teachers' workload—participants identified several notable limitations. These issues primarily relate to the flexibility and granularity of the Individual and Common Characteristic Summarization Modules.
Regarding individual characteristic summarization, the lack of tailored guidelines for capturing personalized traits (e.g., inferring specific preferences from broader categories) constrained the system's reasoning capabilities. This led to an overreliance on explicit configurations rather than flexible inference.
As T7 noted, "\textit{the content fully aligns with the configuration, but inferring specific preferences, such as a child liking 'lions' when configured with 'adventure' and 'animals', would provide more effective guidance}."
Regarding common characteristic summarization, the system struggled to distinguish fine-grained differences within shared attributes. As T5 explained, "\textit{two children may both like sports, but their specific preferences differ; these fine-grained differences within the commonality also need to be reflected}."
These limitations occasionally resulted in misalignments within the three tasks. For example, teachers noted that story frameworks sometimes "\textit{exceed the child's vocabulary level}" (T1); questions "\textit{are not always tailored}" (T3); and explanations "\textit{should use simpler vocabulary}" (T1).}

\textcolor{black}{Moreover, teachers identified areas for improvement in other modules. The Configuration Module was seen as having redundant steps, with T2 noting, "\textit{it could be automatically filled in}," and the granularity of language ability settings was deemed insufficient, with T8 stating, "\textit{vocabulary, grammar comprehension, and expressive abilities are not accurately configured}."
The Review and Feedback Module requires improvements in detail and credibility. T2 suggested that "\textit{it should add more details, such as speaking duration}," while T4 noted that "\textit{speculative AI-generated feedback affected authenticity}." Finally, regarding controllability, T6 mentioned that "\textit{some content could not be modified}," and teachers expressed a desire for the system to "\textit{support teachers with diverse backgrounds}" (T3).}

\textcolor{black}{In conclusion, SparkTales effectively supports coordinators in conducting cross-language collaborative storytelling. By integrating individual and common characteristics, the system generates content for the three crucial tasks that dynamically balances personal traits with shared attributes. This approach improves both coordination efficiency and guidance quality, ultimately reducing the coordinator's workload.}

\subsubsection{\textcolor{black}{RQ2: How can SparkTales help improve children's engagement?}}

\begin{figure*}[t]
    \begin{minipage}[t]{0.24\linewidth}
        \centering
        \includegraphics[width=\textwidth]{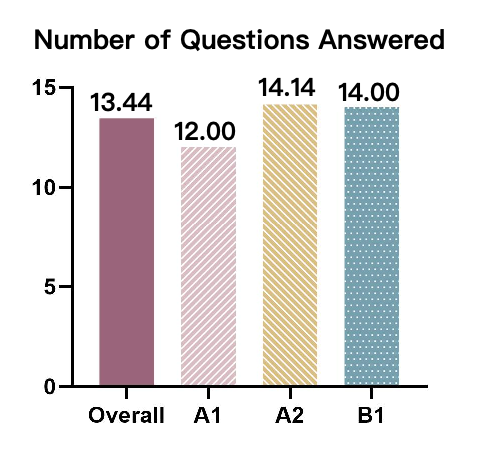}
        \centerline{\small{(a) Number of Questions Answered.}}
    \end{minipage}%
    \hfill
    \begin{minipage}[t]{0.24\linewidth}
        \centering
        \includegraphics[width=\textwidth]{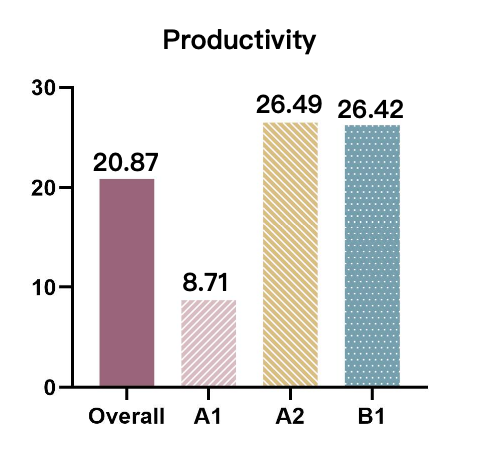}
        \centerline{\small{(b) Productivity.}}
    \end{minipage}
    \hfill
    \begin{minipage}[t]{0.24\linewidth}
        \centering
        \includegraphics[width=\textwidth]{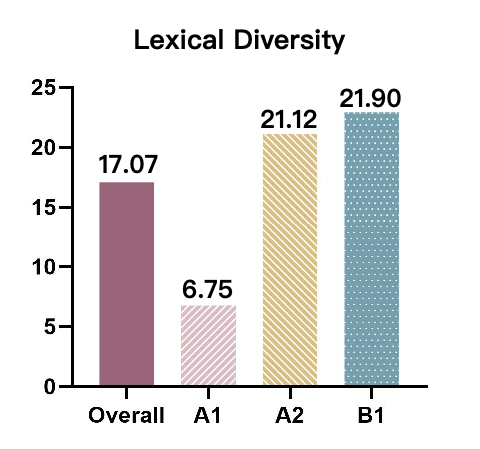}
        \centerline{\small{(c) Lexical Diversity.}}
    \end{minipage}
        \hfill
    \begin{minipage}[t]{0.08\linewidth}
        \centering
        \includegraphics[width=\textwidth]{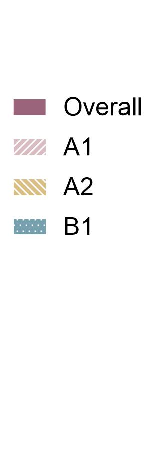}
    \end{minipage}
    \vfill
    \caption{Verbal engagement results for Number of Questions Answered, Productivity and Lexical Diversity.}
    \label{fig: Eva-verbal-1}
\end{figure*}

\begin{figure*}[t]
    \begin{minipage}[t]{0.24\linewidth}
        \centering
        \includegraphics[width=\textwidth]{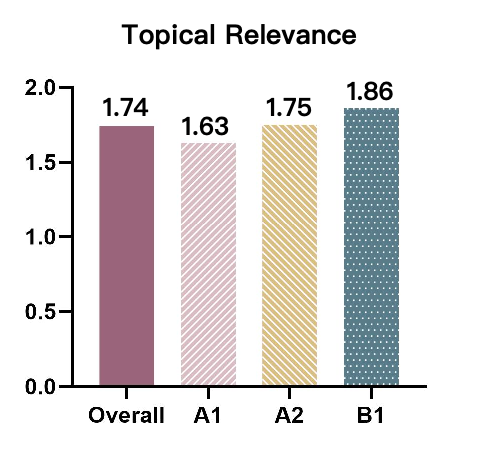}
        \centerline{\small{(a) Topical Relevance.}}
    \end{minipage}%
    \hfill
    \begin{minipage}[t]{0.24\linewidth}
        \centering
        \includegraphics[width=\textwidth]{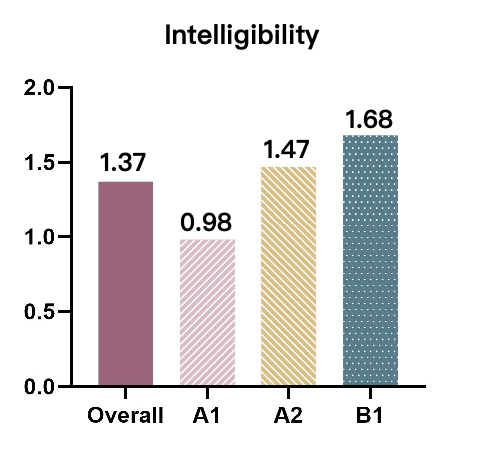}
        \centerline{\small{(b) Intelligibility.}}
    \end{minipage}
    \hfill
    \begin{minipage}[t]{0.24\linewidth}
        \centering
        \includegraphics[width=\textwidth]{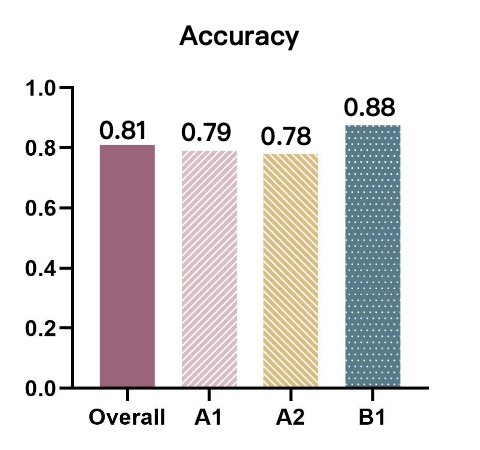}
        \centerline{\small{(c) Accuracy.}}
    \end{minipage}
    \hfill
    \begin{minipage}[t]{0.08\linewidth}
        \centering
        \includegraphics[width=\textwidth]{fig/results/Verbal/legend.png}
    \end{minipage}
    \vfill
    \caption{Verbal engagement results for Topical Relevance, Intelligibility and Accuracy.}
    \label{fig: Eva-verbal-2}
\end{figure*}

We evaluated children's verbal engagement across six dimensions, each including an overall mean value as well as the mean values for each language level involved in this study (A1, A2, and B1) (Figure~\ref{fig: Eva-verbal-1} and Figure~\ref{fig: Eva-verbal-2}). To validate these results, five teachers were randomly selected to rate each dimension's overall value on a five-point Likert scale (5 = highest engagement, 1 = lowest), based on their experience and observations. Overall, SparkTales supports high verbal engagement and expressive ability in children. The overall average number of questions answered was 13.44 (rated 4.0 by teachers), reflecting high response frequency. For language productivity and lexical diversity, overall means were 20.87 and 17.07 (each rated 4.4 by teachers), reflecting active language generating with diverse vocabulary. Topical relevance 1.74/2 (rated 4.0 by teachers) and intelligibility 1.37/2 (rated 3.8 by teachers) show that mostly clear expression, with occasional ambiguity in sentences or pronunciation. Accuracy 0.81/1 (rated 3.8 by teachers) shows that most children responded correctly under coordinator guidance, though some inaccuracies arose from task-language proficiency mismatches. To conclude, across six dimensions and teacher validations, SparkTales enables coordinators to effectively guide children, enhancing verbal engagement and expression, particularly in productivity and lexical diversity.

We further analyzed children's performance across language proficiency. Question responses averaged 12.00 for A1, versus 14.14 and 14.00 for A2 and B1, reflecting participation limits from vocabulary and comprehension. In verbal productivity, A1 children averaged 8.71 words per response, compared with 26.49 for A2 and 26.42 for B1, suggesting higher proficiency enables longer and more coherent sentences; lexical diversity similarly increased with proficiency (A1: 6.75; A2: 21.12; B1: 21.90), reflecting richer vocabulary and greater expressive engagement. Topical relevance scores were 1.63, 1.75, and 1.86, showing that higher-proficiency children more consistently followed activity topics, while lower-proficiency children occasionally deviated. Intelligibility averages were 0.98, 1.47, and 1.68, revealing clearer expression with higher proficiency. Accuracy averaged 0.79, 0.78, and 0.88, suggesting most children completed tasks correctly, although lower-proficiency children were more likely to make errors due to comprehension or expressive limitations. Overall, these results indicate that higher language proficiency corresponds to improvements across dimensions.

\textcolor{black}{In summary, during the collaborative storytelling supported by SparkTales, children demonstrated a significant increase in linguistic engagement. 
Interviews with coordinators further echoed these results. Coordinators noted that the system accurately identified children's individual and common characteristics in the configuration and generated tailored content for the three core tasks, thus capturing children's traits and encouraging their active involvement. This enabled children of varying proficiency levels to participate actively within their comprehension and expression capabilities.
For example, when the story framework featured two children who liked adventure, the teacher noted, "\textit{the children were more engaged and spoke more when the paragraph was related to adventure content}" (T6). Similarly, for a girl who liked Elsa, the teacher noted, "\textit{she was more proactive when the questions involved Elsa}" (T8).
In the material generation phase, another teacher highlighted, "\textit{the generated images and text closely matched the children's interests, sparking their enthusiasm and curiosity}" (T7).}


\section{Discussion}
\label{Discussion}

Our evaluation indicates that SparkTales demonstrates significant potential in supporting cross-language collaborative storytelling. By synthesizing insights from the formative study, design process, and evaluation, the following discussion synthesizes several key insights: \textbf{(1) the transformation of coordinators' roles through Coordinator–AI collaboration}; \textbf{(2) mechanisms for fostering children's engagement via common and individualized features}; \textbf{(3) the balance between coordinator control and AI visibility}; and \textbf{(4) the broader generalizability of these interaction principles in the design of AI-supported educational systems}.

\subsection{Transforming Coordinators' Role in Cross-Language Collaborative Storytelling}

The introduction of SparkTales shifts traditional cross-language collaborative storytelling from coordinator-led to coordinator–chatbot collaboration, significantly reducing the coordinators' workload in handling multiple tasks. However, this shift poses several challenges that necessitate exploration in the future.

In cross-language collaborative storytelling, SparkTales significantly enhances coordinators' efficiency across different stages. During preparation, it simplifies material collection and design for children from diverse language backgrounds, addressing a gap in prior AI-supported collaboration studies that primarily focus on the adaptation to activity organizers rather than participants \cite{ullmann2024towards, shneiderman2020human, workflow}. SparkTales leverages LLMs to summarize children's traits and interests, and generate appropriate story frameworks, ensuring content suitability while reducing coordinators' workload. During storytelling, SparkTales alleviates coordinators' multitasking challenges. While prior studies have explored similar tasks, such as question-answer generation \cite{yao2012semantics, zhao2022educational}, multimodal content generation \cite{shoeibi2023cross}, and real-time feedback \cite{kim2020ai}, these approaches are general and lack suitability for cross-language storytelling interactions. Our study innovatively integrates AI support to address the unique challenges of coordinators in cross-language collaborative storytelling, alleviating their burdens and enabling greater focus on children's interaction. Many teachers highlighted this as the most enjoyable "spark" moments, when they could fully engage their facilitation roles, creating a more dynamic atmosphere. In the review stage, SparkTales automatically generates engagement reports and feature-based feedback, reducing the coordinator's workload in reflection, whereas existing studies have primarily emphasized static reports and visualizations based on uniform metrics \cite{shum2012social, ifenthaler2019utilising, wang2005design}. By integrating children's pre-configured features with story content (e.g., the seven \textit{attribute} story elements), SparkTales generates relevant individualized review and feedback, enhancing the comprehensiveness and immediacy of the review while guiding for subsequent adjustments. Furthermore, SparkTales also reduces coordinators' reliance on knowledge, experience, and skills, enabling novice or less experienced teachers to receive real-time support and gain facilitation, while also allowing other roles, such as parents, to effectively organize and guide cross-language collaborative storytelling, substantially expanding the scope and accessibility.
\textcolor{black}{These findings align with Human–AI Teaming collaboration by positioning AI as an assisting partner that shares cognitive responsibilities with the coordinators rather than replacing them. 
By redistributing routine and analytical tasks, such as story generation, question design, and engagement feedback, between the coordinators and AI, SparkTales reduces coordinators' workload and cognitive pressure, allowing them to facilitate interactions, make adaptive decisions, and support children's engagement.}

However, the introduction of SparkTales can pose challenges for coordinators. First, coordinators with different experiences encounter various usage issues. Less experienced coordinators (e.g., T1, T7) tend to over-rely on LLM-generated content, such as stories and questions, without adapting to children's behaviors or activity context, diminishing their role in cross-language activities \cite{luckin2016intelligence}, whereas experienced coordinators (e.g., T6) critically evaluate system's outputs and spend more time manually adjusting content, increasing operational burden \cite{holmes2019artificial, machado2025workload}. Second, AI's introduction can influence coordinator–child interactions, as coordinators' attention may be distracted from sustaining children's engagement by manipulating AI support, leading to delays or lapses in guiding discussions, responding to needs, and regulating activity flow. Thus, future research could consider more adaptive support strategies tailored to coordinators with different experience levels. For example, using layered guidance and recommendation mechanisms to guide novices while preserving autonomy for experienced coordinators, enhancing the system's adaptability and decision-support capacity. In addition, it is suggested to explore more efficient forms of AI intervention to minimize interference to coordinator–child interactions, ensuring AI's appropriate introduction and usage.

\subsection{Facilitating Children's Engagement through Coordinator-AI Collaboration}

SparkTales leverages similarities and differences in children's features to alleviate language barriers and stimulate expression, significantly enhancing participation in cross-language collaborative storytelling. Nonetheless, designs to balance participation and consider children's features can also bring potential issues. 


SparkTales leverages both common and individual characteristics of children to enhance participation. First, common characteristic utilization is particularly important, as prior studies have shown that children with similar interests and cognitive levels are more likely to establish shared topics, actively expressing ideas and responding to peers \cite{boivin2023co, liu2010children}. SparkTales implements a Guideline-driven mechanism using semantic matching and reasoning to identify shared characteristics, and leverages these commonalities to generate content - such as story frameworks - aligned with both children's profiles, stimulating reciprocal expression and collaborative exploration. Secondly, the system provides personalized content and feedback based on children's individual characteristics. Prior studies indicate that personalized multimodal content can significantly boost children's engagement and language expression \cite{ye2025colin, kewalramani2024systematic, agrawal2022personalized, arifin2025influence}, while continuous personalized feedback further aligns content with children's features to deepen involvement \cite{de2020effects}. In designing SparkTales, we tailor questions and multimodal materials to each child's features, such as language proficiency and preferences, ensuring participation within their ability, while adapting subsequent content and generating feedback from real-time verbal output to encourage narrative exploration and support deeper interaction.

Notably, by integrating common and individual characteristics, SparkTales broadens the applicability of traditional cross-language collaborative storytelling, which typically requires matching children based on similar features, such as language proficiency (e.g., comparable vocabulary size and grammatical expression)  \cite{warschauer2000network, walqui2000context}. On one hand, SparkTales analyzes children's commonalities in vocabulary and grammar to generate content at an appropriate difficulty level, allowing all participants to engage within their abilities - for instance, generating medium-difficulty frameworks that consolidate knowledge for high-proficiency children while introducing new concepts to low-proficiency ones. Regarding individual characteristics, SparkTales tailors content generation to each child's proficiency - using complex structures for advanced learners and simplified or native-language supports for beginners - enabling practice within their comprehension. Thus, shared content builds a common foundation while personalized content addresses individual needs, reducing communication barriers and supporting effective expression, comprehension, and language practice.


Although SparkTales effectively fosters children's participation, it also brings some challenges. First, SparkTales balances speaking opportunities by regulating the number of questions each child answers, a form of "absolute fairness" (i.e., equal distribution of opportunities among participants) \cite{tam2011quantification, tam2014quantification} that prevents highly active children from dominating while ensuring less active children can equally express themselves. However, when activities involve diverse goals or open-ended tasks, such as brainstorming, absolute fairness may suppress children's potential and initiative, limiting natural and creative interactions. In such cases, more flexible balancing strategies could be explored, such as "relative fairness" (i.e., distribution of opportunities relative to one's own performance and that of peers) \cite{tam2011quantification, tam2014quantification}, which maintains basic balance while allowing differentiated allocation. For instance, in vocabulary or grammar exercises, children who need additional practice could receive more speaking turns to reinforce expression, while those who have already mastered the content could get fewer but more refined opportunities, encouraging attentive listening and reflection for more creative and meaningful contributions. Second, conducting personalization for children's learning activities can pose some challenges. Tailoring materials to children's features, such as abilities and interests, may overemphasize familiar domains, limiting exploration of broader content and potentially causing polarization. Coordinators may further aggravate this problem by selecting simple questions to encourage children's engagement, potentially neglecting more challenging tasks that foster language development and deeper thinking. Additionally, there may be some risks to children in AI-generated content, and some of them are difficult to identify by coordinators, such as implicit toxicity \cite{LifeTox}. To conclude, future research could explore flexible balancing mechanisms across activity stages, e.g., applying absolute fairness in structured exercises to ensure participation, and relative fairness in open-ended or creative contexts to allow differentiated engagement. Simultaneously, more indicators should be considered in question and material generations, such as diversity and safety, to ensure children's engagement as well as healthy development.

\subsection{Balancing Control and Visibility within a Triangular Dynamic}

SparkTales introduces advanced AI models (LLMs) to facilitate storytelling while ensuring coordinator control, bringing benefits to both coordinators and children, but also raising issues that warrant further exploration.


In SparkTales, coordinators have flexible control over the AI, which is essential for the system to be supportive rather than substitutive, thereby ensuring safety and trust in the AI-assisted process \cite{kieseberg2023controllable, yampolskiy2022controllability}. To ensure controllability, SparkTales allows coordinators to engage easily - for example, by editing story content, selecting multidimensional questions, and managing multimodal outputs - while simultaneously supporting flexible AI adjustments according to context, such as reducing assistance in familiar areas, thereby ensuring adaptable and context-sensitive support.


However, coordinator control may also introduce issues. The first is over-reliance on or overlooking AI's generations. Over-reliance on coordinators' management of AI outputs may increase cognitive and operational workload, whereas over-reliance on AI-generated content may produce outputs misaligned with storytelling goals or even pose risks to children. 
\textcolor{black}{Additionally, given the two AI–coordinator collaboration modes discussed in Section \ref{Related Works: Coordinator–AI Teaming for Child Interaction}, we design SparkTales as an assistant for coordinators while remaining "unremarkable" \cite{yang2019unremarkable} to children. 
Accordingly, LLMs support the coordinator rather than interacting directly with children, thereby bringing an issue of AI "visibility" (i.e., the extent to which AI's presence and impact can be perceived \cite{shi2023understanding}).} 
Studies have shown that in activities involving adult participants, such as teachers or parents, reducing AI visibility to children helps maintain children's focus and foster trust in adults \cite{escobar2022guidelines, van2023children}. In SparkTales, direct AI interaction with children is minimized by presenting only content chosen by the coordinator, which, as observed by T2, allows children to engage naturally without distractions such as joking or challenging the AI. Simultaneously, coordinators' control over AI outputs safeguards the safety and appropriateness of information for children. However, completely invisible AI may limit children's exposure to new technologies and constrain autonomous exploration. Some teachers (T1, T3) suggested making AI controllably accessible to children, especially older ones, to leverage advanced technology and expand the ways children participate. Future research could explore flexible AI controllability and visibility within the coordinator–child–AI triangular dynamic, such as adaptive coordinator control and child–AI interactions within coordinator-regulated conditions.

\subsection{Generalization}

\textcolor{black}{In this study, we focus on the design and evaluation of SparkTales within a specific cross-cultural collaborative storytelling scenario: a teacher coordinating two bilingual (Chinese–English) children aged 7–11. However, the core mechanisms of SparkTales hold the potential to be generalized to scenarios involving different age groups, languages, and coordinator roles.}

\textcolor{black}{First, regarding age adaptation, the current implementation targets children aged 7–11, tailoring content based on age-appropriate guidelines for language ability and content preferences.
Building on this design, SparkTales could be extended to other age groups.
In the Configuration Module, feature tags could be expanded to accommodate a wider age range (e.g., pre-school or adolescence).
Similarly, the guidelines within the Characteristic Summarization Modules could be replaced with frameworks adapted to the developmental stages of other groups.
For instance, for adolescents aged 12–18, feature modeling could be informed by narrative types found in Young Adult (YA) literature \cite{malo2019critical}.
Furthermore, in the three core tasks—story framework, question, and material generation—prompts could be adjusted to align vocabulary difficulty and interaction style with the target group's cognitive level.
However, extending the system to younger children presents challenges: generated story frameworks and instructions might exceed their cognitive capacities in terms of information density, potentially causing cognitive overload and hindering participation.}

\textcolor{black}{Second, regarding linguistic context, SparkTales can be adapted beyond the Chinese–English context by refining prompt instructions and incorporating themes relevant to the target language.
For example, in a French–American scenario, the Configuration Module could include a "French" language option. Consequently, the Characteristic Summarization Modules would incorporate guidelines based on French children's literature to capture relevant cultural characteristics \cite{frechliterature}.
Additionally, prompts in the content generation modules could be refined to include cultural cues, such as "\textit{generate story paragraphs for French children that include basic English vocabulary... and elements of French rural life}."
However, for minority or low-resource languages, the limited availability of training corpora may constrain the LLM's understanding of vocabulary, grammar, and cultural nuances \cite{grenoble2014minority, valdes1995teaching}.
In such cases, reliance on LLM-generated content may lead to inaccuracies, increasing the coordinator's workload and potentially undermining children's comprehension.}

\textcolor{black}{Finally, while this study evaluates teacher-led coordination, SparkTales could be adapted for contexts where parents serve as coordinators.
Unlike teachers, parents often lack pedagogical expertise and bilingual proficiency, which may increase their reliance on system-generated content.
Limited proficiency in the target language can complicate coordination tasks, such as vocabulary selection and difficulty adjustment. For instance, during the story completion phase, parents might strictly adhere to system-provided questions without the confidence to adapt them.
These challenges increase the burden on parents and may negatively impact the quality of interaction and sustained engagement if the system does not provide sufficient scaffolding.}

\section{Ethical Consideration}
\label{Ethical Consideration}

During SparkTales' development, we prioritized ethical considerations, especially for child participants. First, we submitted a research application to the university's Institutional Review Board (IRB), providing all required materials in accordance with regulations, and obtained application approval. During user research and evaluation, we strictly followed standardized procedures for teacher and child participation. Teachers provided informed consent after being fully briefed on the study's purpose, procedures, and their rights, while for children, we obtained parental written consent alongside the child's verbal assent to ensure voluntary and informed participation. Participants were informed of their right to withdraw at any time, and all data were anonymized, securely stored, and transmitted with encryption to protect privacy and identity.


\section{Limitations and Future Work}
\label{Limitations and Future Work}
As an exploratory study, this research has several limitations that future work should address.

First, this study focuses on a specific paired cross-language collaborative storytelling scenario with fixed participants and tasks, limiting coverage of complex interaction patterns and diverse linguistic needs, which may restrict the system's adaptability to larger, more heterogeneous groups or different learning objectives. Future research could examine its application in multi-party collaboration and non-child populations, such as generating frameworks from existing stories and introducing vocabulary via diverse methods, to enhance generalizability and practical impact.


\textcolor{black}{Second, the current formative study and evaluation mainly focus on teachers from a single country as coordinators, with child participants limited to Chinese and English learners.
The narrow participant pool can limit the understanding of coordinators' strategies and children's interactions across different settings,  constraining the applicability of the findings to more diverse cultural and linguistic contexts. Future research should expand the sample to include coordinators and children from diverse languages, cultures, and roles to examine how these factors influence collaborative storytelling and system use.}



\textcolor{black}{Finally, the evaluations lack long-term assessments, as the current study only examined a short storytelling session rather than sustained use over time. 
As a result, we could not assess the system's long-term impact on coordinators' practices, such as potential reliance on the tool, or on children's collaborative behaviors, such as sustained engagement. This limits our understanding of the system's broader practical value and generalizability.
Future studies should involve longer-term deployments and adopt a broader, multi-dimensional set of assessment measures to systematically evaluate the system, identifying emerging challenges for continuous optimization.}



\section{Conclusion}
\label{Conclusion}
In this study, we developed SparkTales to address the multifaceted tasks and cultural challenges coordinators face in cross-language collaborative storytelling, helping to foster deeper interaction and active participation among children. Through a formative study, we identified coordinators' needs and expectations, which informed the design of SparkTales with multi-stage AI-assisted functions. Real-world evaluations demonstrated SparkTales' effectiveness in reducing coordinators' workload and promoting children's engagement, while also revealing some limitations in personalization and diversity of user experience. These findings inspire future work to develop adaptive coordinator-oriented strategies and dynamic interaction mechanisms to balance children, coordinators, and AI, while enhancing the system's applicability and generalizability.

\section{Acknowledgments of the Use of AI}
\label{Acknowledgments of the Use of AI}
We used AI, specifically LLMs, to generate content for cross-language collaborative storytelling, including story frameworks, diverse questions, and multimodal materials. Detailed descriptions of AI usage are provided in Section \ref{SparkTales}, with prompts provided in Section \ref{Appendix_Prompts}. The authors take full responsibility for the output and use of AI in this paper.

\begin{acks}

This work is supported by National Key Research and Development Program of China under the Grant No. 2024YFC3307401 and Major Project of the National Social Science Fund of China (NSSFC) under the Grant No. 25\&ZD260. Peng Zhang is a faculty of College of Computer Science and Artificial Intelligence, Fudan University. Tun Lu is a faculty of College of Computer Science and Artificial Intelligence, Shanghai Key Laboratory of Data Science, and MOE Laboratory for National Development and Intelligent Governance, Fudan University.

\end{acks}

\bibliographystyle{ACM-Reference-Format}
\bibliography{main}

\appendix

\section{Appendix}
\subsection{SparkTales' Prompts}
\label{Appendix_Prompts}

\begin{figure*}
    \begin{minipage}[t]{0.25\linewidth}
        \centering
        \includegraphics[width=\textwidth]{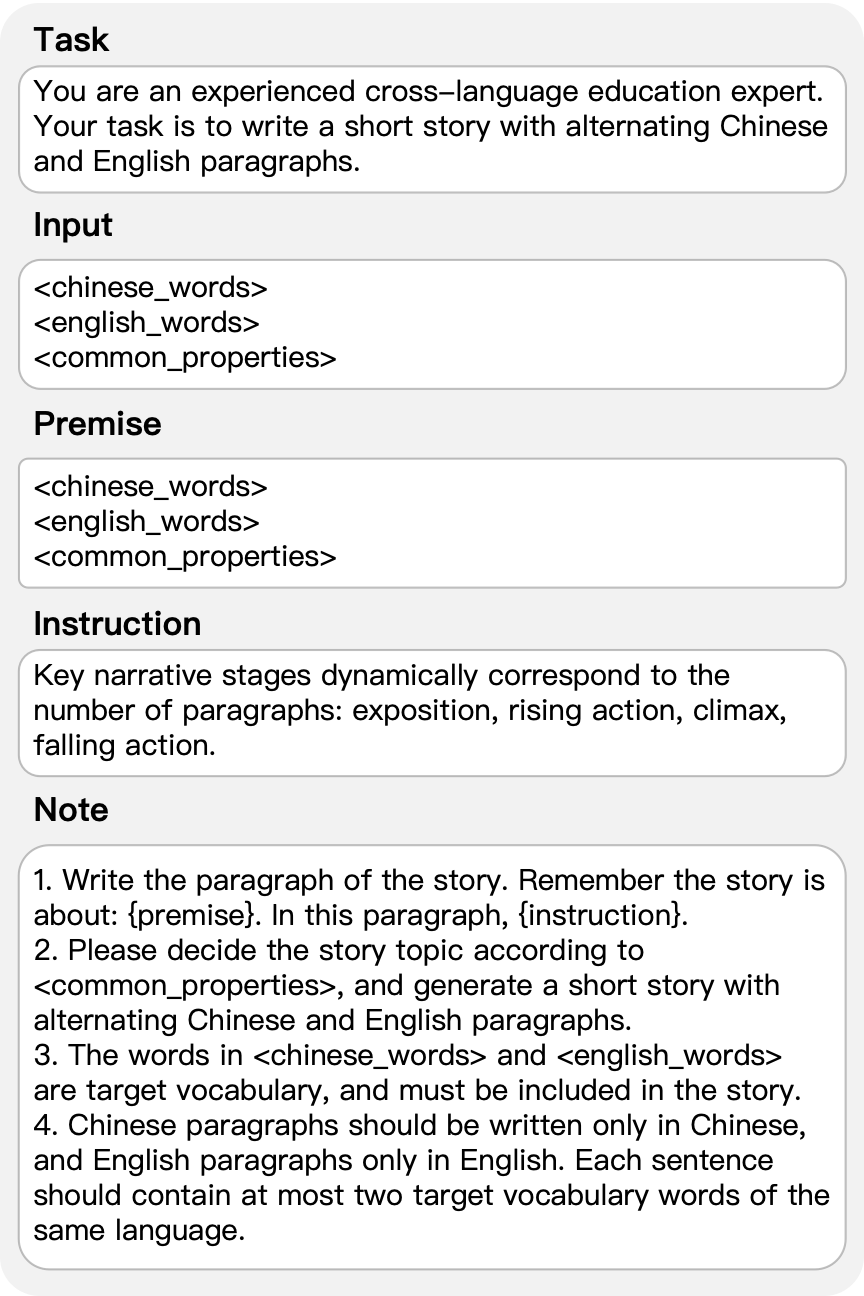}
        \centerline{\small{(a) Prompt for story generation.}}
    \end{minipage}%
    \hfill
    \begin{minipage}[t]{0.25\linewidth}
        \centering
        \includegraphics[width=\textwidth]{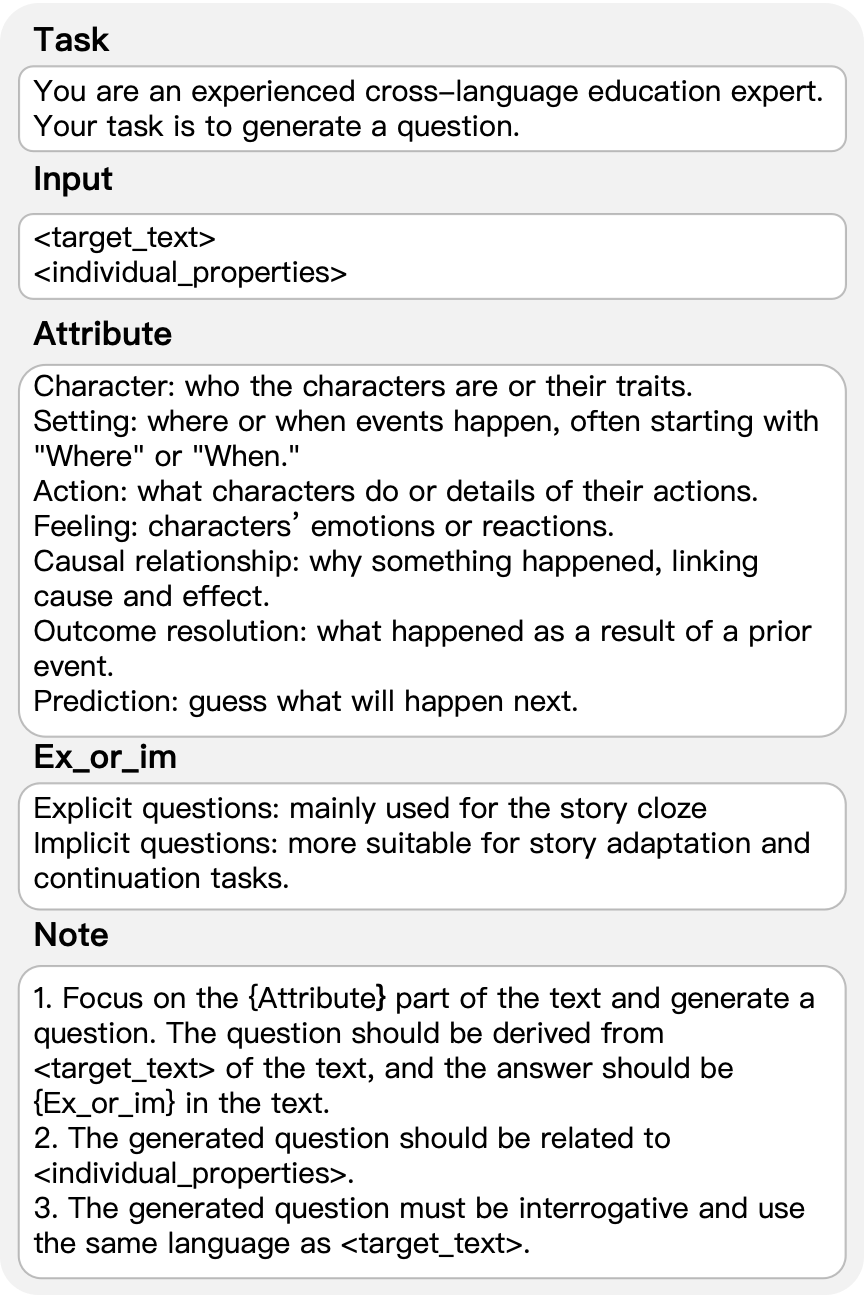}
        \centerline{\small{(b) Prompt for question generation.}}
    \end{minipage}
    \hfill
    \begin{minipage}[t]{0.25\linewidth}
        \centering
        \includegraphics[width=\textwidth]{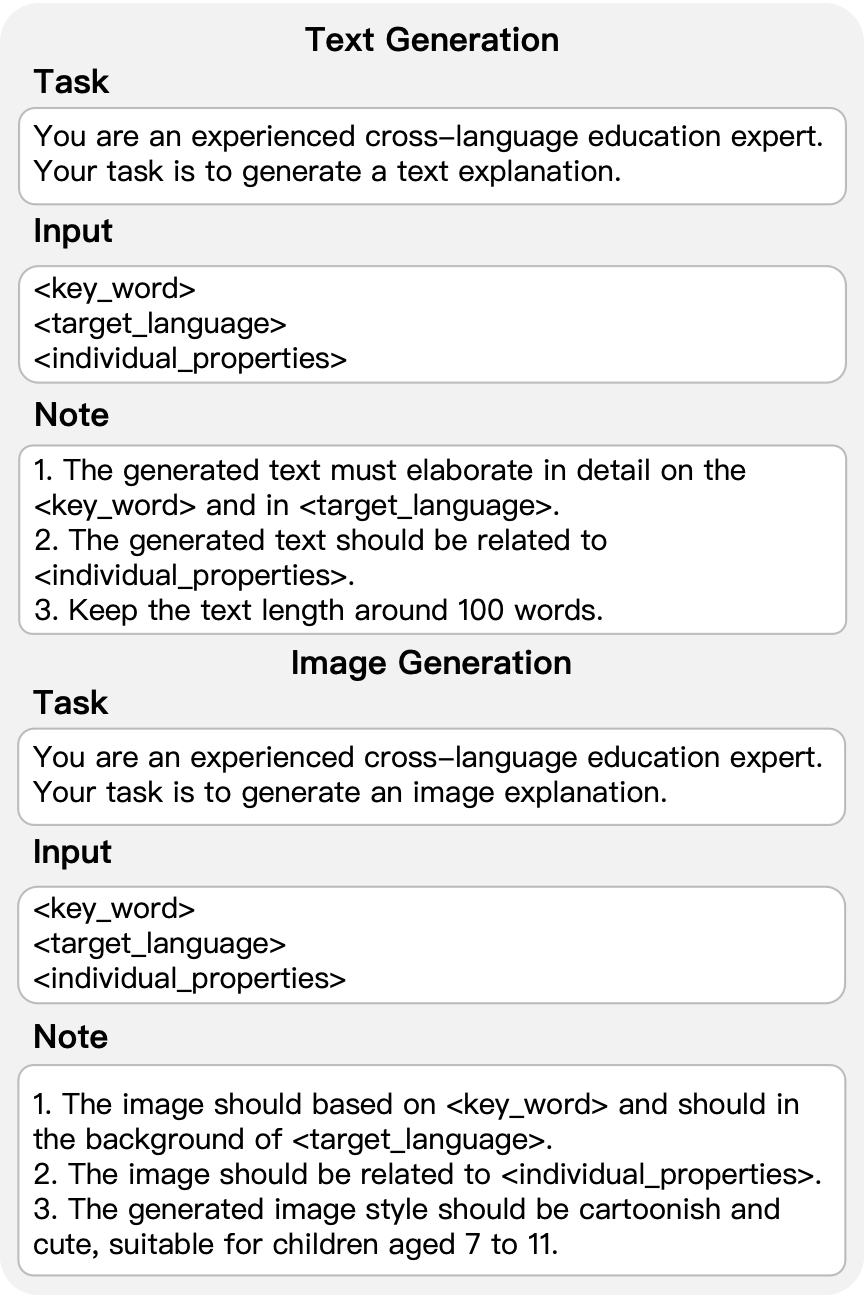}
        \centerline{\small{(c) Prompt for material generation.}}
    \end{minipage}
    \vfill
    \caption{Prompts for SparkTales.}
    \label{fig: prompt}
\end{figure*}

The prompts of SparkTales are illustrated in Figure~\ref{fig: prompt}, where (a), (b) and (c) present the prompts for story generation, question generation, and comprehension-oriented material generation, respectively.

\end{document}